\documentclass[journal]{IEEEtran}

\usepackage{tikz}
\usepackage{amsmath}
\usepackage{comment}
\usepackage{color}
\usepackage{booktabs}
\usepackage{xspace}
\usepackage{amssymb}
\usepackage{amsfonts}           
\usepackage{mathrsfs} 
\usepackage{multirow}
\usepackage{booktabs}
\usepackage{graphicx}
\usepackage{enumitem}
\usepackage{mathtools}
\usepackage{cite}
\usepackage{caption}
\usepackage{subcaption}
\usepackage{tabularx}
\usepackage[multiple]{footmisc}

\newcommand{\maRevise}[1]{\textcolor[rgb]{0,0,0} {#1}}

\begin{document}
%
\title{Stepwise-Refining Speech Separation Network via Fine-Grained Encoding in High-order Latent Domain}
%
%
%

\author{
Zengwei Yao,
Wenjie Pei$^\star$,
Fanglin Chen,~\IEEEmembership{Member,~IEEE,}
Guangming Lu$^\star$,~\IEEEmembership{Member,~IEEE,}
and David Zhang,~\IEEEmembership{Life~Fellow,~IEEE}
\thanks{$^\star$Wenjie Pei and Guangming Lu are corresponding authors.}
\thanks{Zengwei Yao, Wenjie Pei, Fanglin Chen and Guangming Lu are with the Department of Computer Science, Harbin Institute of Technology at Shenzhen, Shenzhen 518057, China (e-mail: yaozengwei@outlook.com; wenjiecoder@outlook.com; chenfanglin@hit.edu.cn; luguangm@hit.edu.cn)}
\thanks{David Zhang is with the School of Science and Engineering, The Chinese University of Hong Kong at Shenzhen, Shenzhen 518172, China (e-mail: davidzhang@cuhk.edu.cn)}

}

\maketitle

\begin{abstract}
The crux of single-channel speech separation is how to encode the mixture of signals into such a latent embedding space that the signals from different speakers can be precisely separated. Existing methods for speech separation either transform the speech signals into frequency domain to perform separation or seek to learn a separable embedding space by constructing a latent domain based on convolutional filters. While the latter type of methods learning an embedding space achieves substantial improvement for speech separation, we argue that the embedding space defined by only one latent domain does not suffice to provide a thoroughly separable encoding space for speech separation. In this paper, we propose the Stepwise-Refining Speech Separation Network (\emph{SRSSN}), which follows a coarse-to-fine separation framework. It first learns a 1-order latent domain to define an encoding space and thereby performs a rough separation in the coarse phase. Then the proposed \emph{SRSSN} learns a new latent domain along each basis function of the existing latent domain to obtain a high-order latent domain in the refining phase, which enables our model to perform a refining separation to achieve a more precise speech separation. 
We demonstrate the effectiveness of our \emph{SRSSN} by conducting extensive experiments, including speech separation in a clean (noise-free) setting on WSJ0-2\maRevise{/3}mix datasets as well as in noisy/reverberant settings on WHAM!/WHAMR! datasets. Furthermore, we also perform experiments of speech recognition on separated speech signals by our model to evaluate the performance of speech separation indirectly.

\end{abstract}

\begin{IEEEkeywords}
Speech separation, high-order latent domain, coarse-to-fine, end-to-end
\end{IEEEkeywords}

%
\IEEEpeerreviewmaketitle

\section{Introduction}
\IEEEPARstart{S}{peech} separation aims to separate out the clean speech signals for each involved speaker from a mixture of speech signals. It plays an important role in speech processing~\cite{Overview}, especially in the scenario of mixed and noisy speech environment. 
Speech separation, particularly under the single-channel condition, is still a highly challenging research problem due to the difficulty of encoding the mixed speech signal into an entirely separable embedding feature space. This paper focuses on single-channel speech separation.

A classical type of methods~\cite{DPCL,DPCL2,PIT,UPIT,ADANet,Chimera++,CASA} for single-channel speech separation is to transform the input mixture of temporal speech signals into the frequency domain employing the Short-Time Fourier Transform (STFT)~\cite{STFT} and then perform speech separation in the frequency domain. 
Whilst this type of methods achieves great improvement on speech separation due to relatively mature techniques on time-frequency transformation and signal processing in frequency domain, it suffers from two limitations for speech separation: 1) most existing methods focus on reconstructing the magnitude of the signal in the frequency domain but ignore the modeling of the phase (which is the other crucial physical property of frequency signals) since there is no sufficiently effective way of modeling the phase yet; 2) performing speech separation in the frequency domain is an effective but not necessarily the optimal way, and it is still doubtful whether the frequency domain is able to provide entirely separable space for speech signals~\cite{Conv-tasnet}.

With great success of deep learning in many fields such as computer vision and machine learning by the powerful capability of feature representation, another research line of speech separation~\cite{TASNET,BLSTM-TasNet,Conv-tasnet, furcanext, latent, DPRNN, MULCAT, sudo, DPTNet, SepFormer, wavesplit} seeks to leverage deep convolutional neural networks (CNNs) to learn an embedding space that is separable for speech signals between different speakers. A remarkable benefit of such methods is that the whole separation procedures including encoding, separation and decoding can be integrated into an end-to-end model, which is in contrast to the sequentially individual steps in aforementioned STFT-based methods. A prominent example of such method is Time-domain Audio Separation Network (TASNET)~\cite{TASNET}, which employs a 1-D convolutional network consisting of multiple convolutional filters to transform the temporal signals within a time slot into a latent embedding space. This latent space can be considered as a 1-order latent domain with the convolutional filters as its basis functions. The input signals are first encoded into such embedding space defined by this latent domain and then the separation and decoding are performed subsequently in this embedding space. Many follow-up works~\cite{BLSTM-TasNet, Conv-tasnet, furcanext, DPRNN, sudo, DPTNet, SepFormer} focus on building better separators upon TASNET to further improve the performance of speech separation.
While such TASNET-based methods have boosted the performance of speech separation substantially, we argue that whether the embedding space defined by only 1-order latent domain suffices to provide a thoroughly separable feature space, particularly in challenging scenarios where speakers with similar speech characteristics are involved in. 
An empirical investigation is conducted by ablation study both quantitatively and qualitatively in Section~\ref{sssec:ablation} and the experimental results validate our doubt.

In this paper, we propose the Stepwise-Refining Speech Separation Network (\emph{SRSSN}), which performs speech separation in a stepwise manner following a coarse-to-fine framework. In the coarse phase, it first conducts a rough separation in a coarse embedding space defined in a 1-order latent domain, which is similar as the typical way of TASNET-based methods~\cite{TASNET,BLSTM-TasNet,Conv-tasnet, furcanext, DPRNN, MULCAT, sudo, DPTNet, SepFormer}. In the refining phase, our \emph{SRSSN} learns a new latent domain along the basis functions of the existing domain in the coarse phase to form a high-order domain space. Then the coarse embedding space is further decomposed into a fine-grained embedding space defined by the constructed high-order domain. As a result, our proposed \emph{SRSSN} is able to re-code the coarsely separated features in the fine-grained embedding space and achieve a more precise separation. To conclude, our \emph{SRSSN} benefits from following advantages:
\begin{itemize}[leftmargin=*]
    \item We design a coarse-to-fine separation framework, which first conducts a rough speech separation and then performs the refining separation in the constructed fine-grained embedding space to achieve more precise separated results.
    \item A Fine-grained Encoding Mechanism is designed specifically to construct a fine-grained embedding space by learning a high-order latent domain space, which enables our \emph{SRSSN} to refine the separation results.
    \item Our proposed model can be readily integrated into any of existing TASNET-based frameworks following the encoding-separation-decoding paradigm. In particular, we investigate the performance of our model by integrating it into two typical separator structures: DPRNN-TASNET~\cite{DPRNN} based on RNN and DPTNET-TASNET~\cite{DPTNet} based on Transformer~\cite{transformer}.
    \item Extensive experiments validate the superiority of encoding in the learned high-order latent domain over in a 1-order latent domain (as most existing methods do). It is demonstrated that our \emph{SRSSN} compares favorably against state-of-the-art methods for speech separation in both the clean (noise-free) setting on WSJ0-2\maRevise{/3}mix dataset and noisy/reverberant settings on WHAM!/WHAMR! datasets. Besides, we also perform experiments of speech recognition as an indirect evaluation way, in which a same automatic speech recognition (ASR) model achieves better performance on the separated speech signals by our model than on the separated signals by other models for speech separation.
\end{itemize}

\section{Related Work}
\noindent\textbf{Frequency-domain-based methods.} 
Frequency domain-based methods transform the mixed speech signal into frequency domain using the STFT~\cite{STFT}. They perform separation in frequency domain and transform the separated representations back into speech signals employing the Inverse Short-time Fourier Transform (iSTFT)~\cite{istft}. 
Hershey et al.~\cite{DPCL,DPCL2} devise a Deep Clustering (DPCL) method which learns a speaker-discriminative vector for each time-frequency bin of the mixture spectrogram and employs clustering to obtain speaker assignments for each time-frequency bin.
Kolbæk et al.~\cite{PIT, UPIT} propose a Permutation Invariant Training (PIT) approach to align multiple estimates for different target speakers, which enumerates all source permutations and uses the permutation with the minimum error to update the network parameters. Wang et al.~\cite{Chimera++} integrate the PIT-based mask-inference network~\cite{UPIT} and DPCL~\cite{DPCL2} regularizer into a multi-task learning framework for better separation performance. 
Inspired by the Computational Auditory Scene Analysis (CASA)~\cite{tra_CASA}, Liu et al.~\cite{CASA} propose a Deep CASA approach, which first estimated spectra for each speaker at each frame and then groups the estimated frame-level spectra to different speakers employing clustering. 

\noindent\textbf{Learnable-latent-domain-based methods.} Learnable latent domain-based methods leverage CNN layers to learn a latent domain for speech separation. 
Luo et al.~\cite{TASNET} first propose the TASNET, which takes speech signal as input and reconstructs the separated sources in an end-to-end manner. 
Most variants of TASNET focus on building effective separation network to model the extremely long encoded sequence and thus to perform a precise representation separation. 
Luo et al.~\cite{Conv-tasnet} propose a fully-convolutional TASNET (Conv-TASNET), consisting of stacking 1-D dilated temporal convolutional layers similar to Wavenet~\cite{wavenet}. 
Shi et al.~\cite{furcanext} devise an improved version of Conv-TASNET equipped with dynamic gated mechanism. 
Tzinis et al.~\cite{sudo} propose a computationally efficient backbone structure similar to U-Net~\cite{unet}, which extracts multi-resolution temporal features through successive downsampling and upsampling. 
Luo et al.~\cite{DPRNN} develop a Dual-Path Recurrent Neural Network (DPRNN), which performs local and global temporal modeling alternately, to capture long-time dependency over entire sequence instead of fixed receptive fields as in Conv-TASNET~\cite{Conv-tasnet}. 
Chen et al.~\cite{DPTNet} devise a Dual-Path Transformer Network (DPTNET), which also adopt the dual-path strategy to handle the long sequence as in DPRNN~\cite{DPRNN} and utilize improved transformer layers incorporating RNN layers to model the context in an order-aware manner. 
Subakan et al.~\cite{SepFormer} propose the Sepformer, 
which is an extension of DPTNET~\cite{DPTNet} by employing deeper and wider Transformer layers, achieves better performance than DPTNET at the cost of much larger model size.
    
Meanwhile, several studies explore to leverage the discriminative speaker representation to improve separation performance. 
Nachmani et al.~\cite{MULCAT} introduce a task to minimize the distance between the speaker embeddings of the estimated signal and target signal, where the speaker embeddings are extracted from a separately trained speaker identification network. 
Zeghidour et al.~\cite{wavesplit} propose the Wavesplit network consisting of a speaker stack and a separation stack, where the speaker stack extracts frame-level speaker-discriminative vectors and obtain speaker centroids employing clustering, and the separation stack estimates isolated speech signals conditioned the speaker centroids.

\begin{figure*}
	\begin{center}
    \includegraphics[width=1.0 \textwidth]{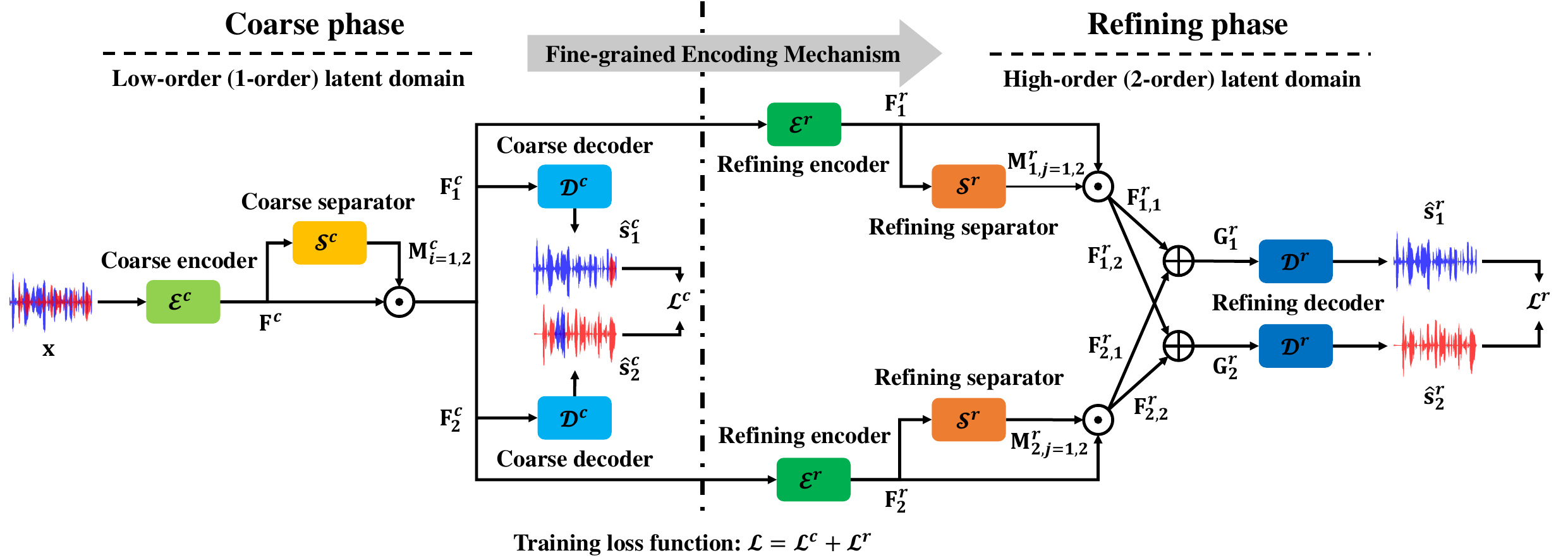}
	\end{center}
	  \caption{ Architecture of our proposed \emph{SRSSN}. It first performs a rough separation in the coarse phase, then the coarsely separated features are further encoded in a fine-grained embedding space to perform a more precise separation in the refining phase. The Fine-grained Encoding Mechanism is specifically designed to learn the fine-grained encoding space for refining separation by constructing a high-order latent domain upon the 1-order latent domain. 
  }
	\label{fig:SRSSN_framework}
\end{figure*}

\noindent\textbf{Two-stage or cascaded architectures.} 
Some studies adopt two-stage or cascaded strategies for speech separation and enhancement. 
Zhao et al.~\cite{two_stage_enc} introduce a two-stage framework to enhance noisy-reverberant speech, where denoising and dereverberation are performed sequentially. 
Kavalerov et al.~\cite{universal} devise a iterative version of TASNET cascading of two same separation model, where the initial estimates obtained in the first model along with the mixture are fed into the second model for a more precise separation. 
Fan et al.~\cite{post} propose to separate the mixture preliminarily in frequency domain and design an End-to-End Post-Filter (E2EPF) to leverage the similarity between the mixture and preliminary estimates to further improve the separation performance.
Delfarah et al.~\cite{two_stage_sep_ehc} introduce a two-stage deep CASA~\cite{CASA} method to separate mixed speech in reverberant condition, where the first stage estimates reverberant separated speech and the second stage dereverberates the separated speech to obtain clean anechoic speech. 
Phan et al.~\cite{GANs} utilize chained generators to gradually refine noisy speech in a stage-wise manner. 


\label{sec:related_work}

\section{Stepwise-Refining Speech Separation Network}
Given a mixture of single-channel speech signal from different speakers, we aim to separate and extract the clean signal for each involved speaker. The crux of this problem is how to encode the mixture signal into such a latent embedding space that different speech sources are entirely separable. To surmount this crux, our proposed Stepwise-Refining Speech Separation Network (\emph{SRSSN}) performs fine-grained encoding in a high-order latent domain rather than in a 1-order latent domain (as most existing methods do) for signals in each time slot, to obtain more precise separation.

Specifically, our \emph{SRSSN} performs speech separation in a stepwise-refining manner following a coarse-to-fine framework. It first conducts a rough separation in a coarse embedding space defined in a low-order (1-order) latent domain, then the coarse embedding space is further decomposed into a fine-grained embedding space defined in a high-order latent domain, which is achieved based on our designed Fine-grained Encoding Mechanism. As a result, the proposed \emph{SRSSN} is able to perform a more precise speech separation. Figure~\ref{fig:SRSSN_framework} illustrates the overall architecture of \emph{SRSSN}. 
We will first present the coarse-to-fine separation framework, then we will elaborate on the Fine-grained Encoding Mechanism to explain how to construct the high-order latent domain for refining the result of speech separation. 
\subsection{Coarse-to-fine Separation Framework}
\label{ssec:coarse-to-fine}
Our \emph{SRSSN} consists of two separation phases: 1) the coarse phase for a rough separation of signals among speakers in a coarse embedding space defined by a 1-order latent domain, and 2) the refining phase in which a fine-grained embedding space is constructed in a high-order latent domain for a more precise separation.

\subsubsection{Coarse phase} As shown in Figure~\ref{fig:SRSSN_framework}, we build both the coarse and refining phases of our model based on the basic encoder-separator-decoder framework adopted by the classical TASNET-based models~\cite{BLSTM-TasNet, Conv-tasnet, furcanext, DPRNN, sudo, DPTNet, SepFormer}. Formally, given a mixture of signal $\mathbf{x}\in \mathbb{R}^{1\times T}$ of temporal length $T$, the coarse encoder $\mathcal{E}^c$ of \emph{SRSSN} encodes the signal 
into a coarse embedding space with $N^c$ basis functions by a nonlinear transformation:
\begin{equation}
\mathbf{F}^c = \mathcal{E}^c (\mathbf{x}).
\end{equation}
Herein $\mathbf{F}^c \in \mathbb{R}^{N^c \times T'}$ is the encoded feature representation and $T'$ is the encoded temporal length. Following the typical encoding scheme~\cite{DPTNet}, we model the coarse encoder $\mathcal{E}^c$ by a 1-D convolutional layer with $N^c$ filters of kernel size $1\times K^c$, followed with the nonlinear activation layer \emph{ReLU}. Here the $N^c$ convolutional filters can be viewed as the basis functions to form a latent domain, which presumably simulates the frequency domain due to the similar mathematical transformations between the encoder and the Short-Time Fourier Transformation~\cite{STFT}. 

The encoded representations $\mathbf{F}^c$ are then fed into the coarse separator $\mathcal{S}^c$ of $\emph{SRSSN}$ to estimate the signal mask $\mathbf{M}_i^c \in \mathbb{R}^{N^c \times T'}, i=1, \dots, D$ for each speaker, where $D$ is the number of involved speakers in the input mixed signal $\mathbf{x}$. The mask values are constrained to be non-negative to indicate the content proportion for each speaker with respect to each basis function by applying a non-linear activation layer $\emph{ReLU}$. Thus the separation in the coarse phase is performed as:
\begin{equation}
    \begin{split}
    & \mathbf{M}_1^c, \dots, \mathbf{M}_D^c = \mathcal{S}^c (\mathbf{F}^c),\\
    & \mathbf{F}^c_i = \mathbf{F}^c \odot \mathbf{M}_i^c, \ \ i = 1, \dots, D,
    \end{split}
\end{equation}
where $\mathbf{F}^c_i \in \mathbb{R}^{N^c \times T'}$ is the separated features for $i$-th speaker and $\odot$ denotes element-wise multiplication. 

\begin{figure}
	\begin{center}
    \includegraphics[width=0.4\textwidth]{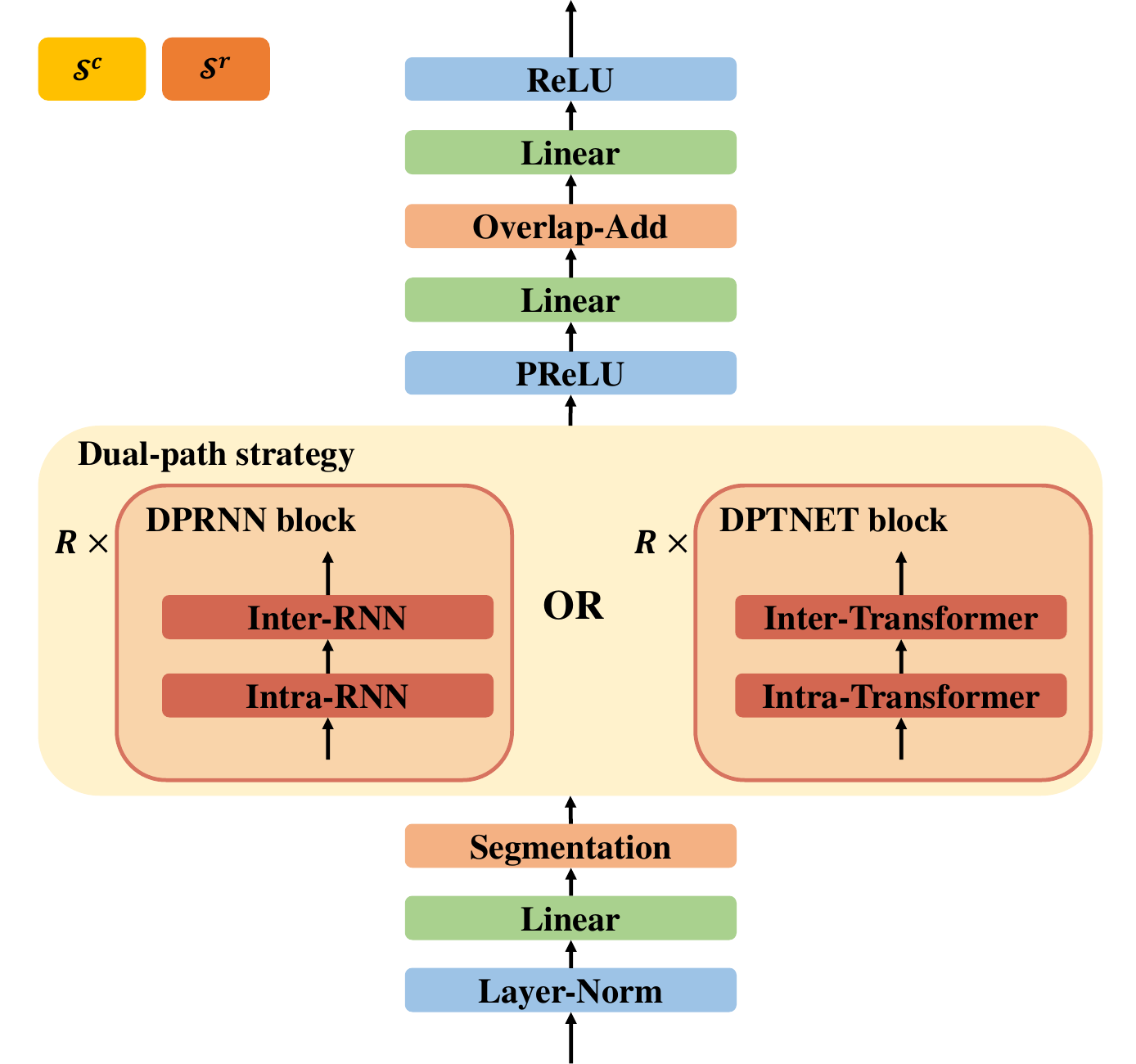}
	\end{center}
	  \caption{Structures of the separator in DPRNN-TASNET and DPTNET-TASNET. Note that the coarse separator $\mathcal{S}^c$ and the the refining separator $\mathcal{S}^r$ share the same model structure.}
	\label{fig:separator}
\end{figure}

\smallskip\noindent\textbf{Modeling of Separator $\mathcal{S}^c$.} The separator $\mathcal{S}^c$ can be implemented in the same structure with any separator of existing TASNET-based models~\cite{BLSTM-TasNet, Conv-tasnet, furcanext, sudo, DPRNN, DPTNet, SepFormer}. To investigate the effectiveness of our model extensively, we evaluate our model equipped with two typical separator structures respectively: the separator structure in DPRNN-TASNET~\cite{DPRNN} and DPTNET-TASNET~\cite{DPTNet} due to their excellent performance. 

Both of the separator structures employ a same core block iteratively (for $R$ times) for modeling temporal dependencies in a speech sequence. The main difference between these two types of separators lies in the structure of the core block: the separator in DPRNN-TASNET employs Bi-directional LSTM~\cite{lstm} to construct its core block (DPRNN block), while the core block (named DPTNET block) of the separator in DPTNET-TASNET is designed based on Transformer~\cite{transformer}. Specifically, as shown in Figure~\ref{fig:separator}, the encoded representation $\mathbf{F}^c$ is firstly normalized by Layer-Normalization~\cite{layernorm} and fed into a linear layer to adjust the number of channels. Then it was equally segmented into successive chunks with temporal length of $L$ and overlap of $\frac{L}{2}$ between adjacent chunks. Both the DPRNN block and the DPTNET block adopt a dual-path strategy to capture the long-term temporal dependencies. Each block consists of two paths: an intra-path used for capturing the intra-dependencies within each chunk and an inter-path for modeling the inter-dependencies among chunks. 
After temporal modeling by $R$ core blocks, a nonlinear activation layer \emph{PReLU}~\cite{prelu} and a linear layer are used to expand the number of channels by $D$ times to make the separated encoding representation $\mathbf{F}^c_i, i=1, \dots, D$ keep the same feature dimension as the un-separated representation $\mathbf{F}^c$. Subsequently, the chunks are transformed back into sequential shape through the overlap-add operation\cite{DPRNN}. Finally, a linear layer followed by a activation layer \emph{ReLU} is utilized to estimate the mask values.

The separated representations $\mathbf{F}^c_i$ are finally decoded into the speech sources $\hat{\mathbf{s}}^c_i \in \mathbb{R}^{1\times T}, i= 1, \dots, D$ for each speaker by the coarse decoder $\mathcal{D}^c$:
\begin{equation}
    \hat{\mathbf{s}}^c_i = \mathcal{D}^c (\mathbf{F}^c_i), i = 1, \dots, D.
\end{equation}
We model the decoder $\mathcal{D}^c$ as a 1-D transposed convolutional layer which is the reversed operation of the encoder. Note that the decoding operation in the coarse phase is only performed in training for supervision. During inference, the separated latent representations $\mathbf{F}^c_i$ are fed into the refining phase for the fine-grained separation.

\begin{figure*}[!t]
\centering
    \begin{minipage}{.50\linewidth}
        \begin{subfigure}[t]{\linewidth}
          \includegraphics[width=\textwidth]{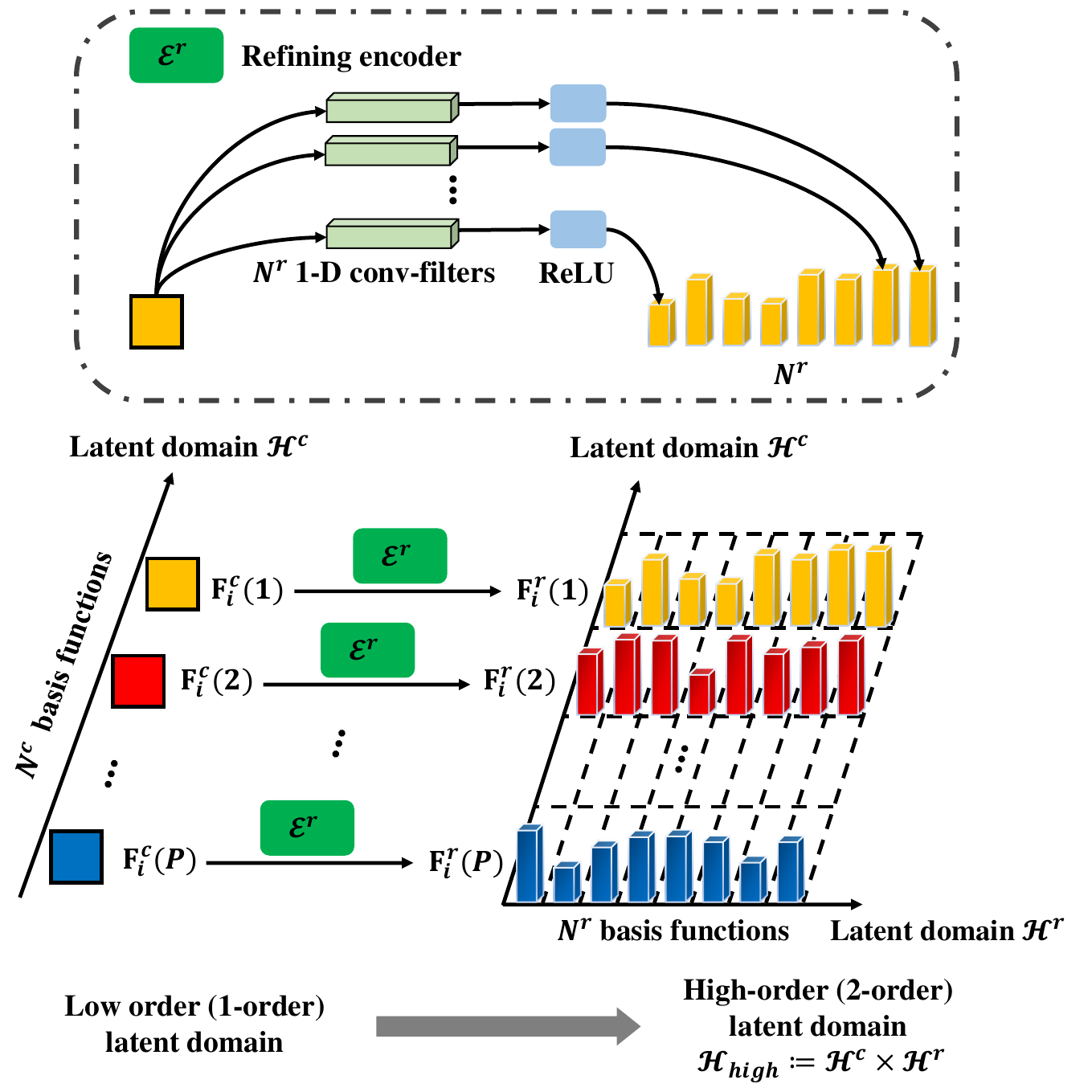}
            \caption{Fine-grained encoding mechanism}
        \end{subfigure}
    \end{minipage}
    \begin{minipage}{.275\linewidth}
        \begin{subfigure}[t]{\linewidth}
            \includegraphics[width=\textwidth]{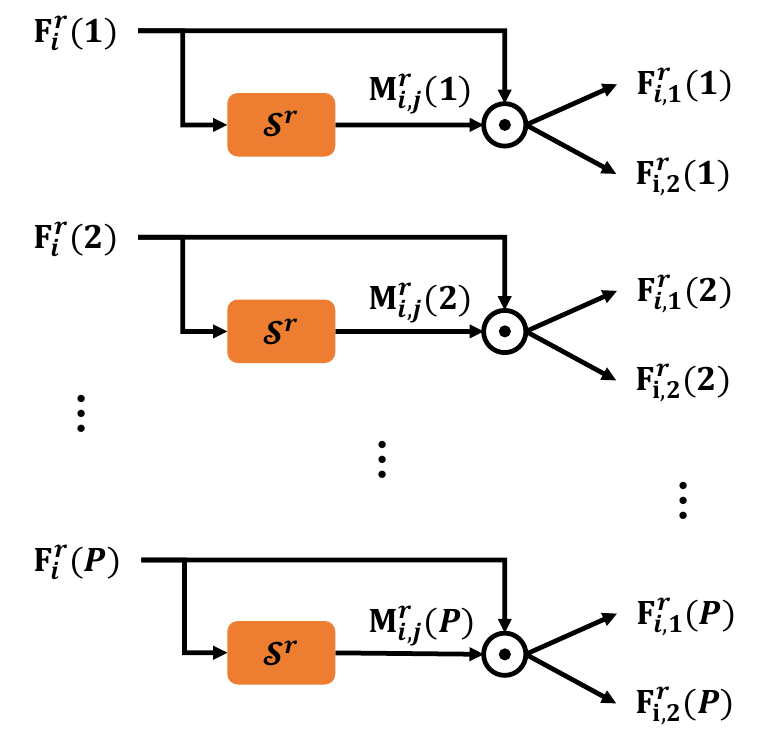}
            \caption{Refining separator $\mathcal{S}^r$}
        \end{subfigure} 
        \begin{subfigure}[b]{\linewidth}
            \includegraphics[width=\textwidth]{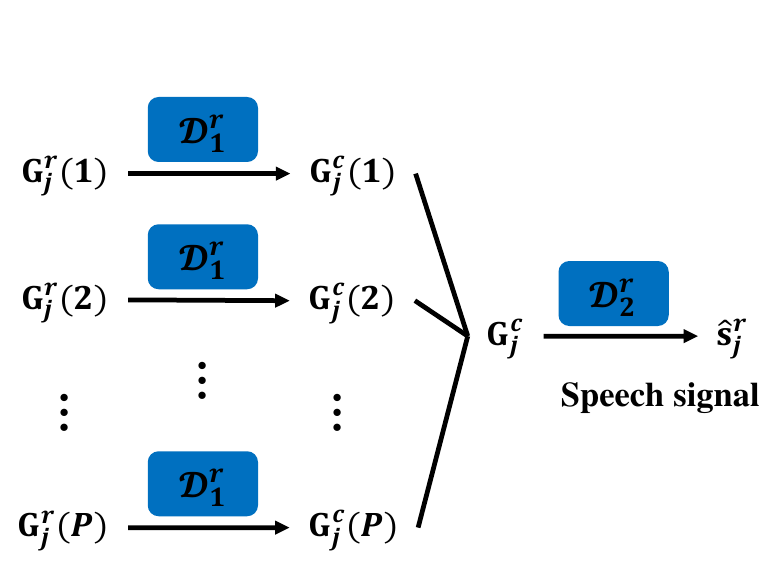}
            \caption{Refining decoder $\mathcal{D}^r$}
        \end{subfigure} 
    \end{minipage}
  \caption{Fine-grained separation in the refining phase. (a) The proposed Fine-grained Encoding Mechanism learns a new latent domain $\mathcal{H}^r$ along each basis function of the old latent domain $\mathcal{H}^c$ that defines the coarse embedding space. The obtained new latent domain $\mathcal{H}^r$ and the old latent domain $\mathcal{H}^c$ jointly form a high-order domain, which defines a fine-grained embedding space for refining separation. (b) The refining separator processes the encoded representations in parallel for all groups of basis functions of the latent domain $\mathcal{H}^c$ for each speaker. (c) The refining decoder performs two-stage decoding to obtain final speech signal: decoding from fine-grained embedding space to the coarse embedding space by $\mathcal{D}^r_1$ and decoding from the coarse embedding space to the output speech signal by $\mathcal{D}^r_2$.}
  
  \label{fig:refining}
\end{figure*}

\subsubsection{Refining phase}
The performance of the whole model relies on the effectiveness of feature separation in the encoding embedding space. Thus, it is crucial that the speech sources can be precisely separable between different speakers in the encoded embedding space. 
However, we argue that the coarse embedding space encoded by only one latent domain does not suffice to provide a thoroughly separable feature space for all speakers, which is validated empirically by the ablation experiments in Section~\ref{sssec:ablation}. To tackle this limitation, we propose the Fine-grained Encoding Mechanism (presented in Section~\ref{ssec:FGEM}) to construct a high-order latent domain and thereby define a fine-grained embedding space upon the coarse space to make a more precise separation.

Taken the separated representations in the coarse embedding space $\mathbf{F}^c$ as inputs, the encoder $\mathcal{E}^r$ in the refining phase re-codes the features for each speaker in the fine-grained embedding space:
\begin{equation}
    \mathbf{F}^r_i = \mathcal{E}^r (\mathbf{F}^c_i), i = 1, \dots, D,
\end{equation}
where the refining encoder $\mathcal{E}^r$ will be described concretely in Section ~\ref{ssec:FGEM}. 

Benefitting from the fine-grained feature decomposition from $\mathbf{F}^c$ to $\mathbf{F}^r$, a refining separation can be performed on $\mathbf{F}^r$ for each speaker by the separator $\mathcal{S}^r$ in the refining phase, which is similar to the separation procedure in the coarse phase. For instance, the encoded feature for the $i$-th speaker $\mathbf{F}^r_i$ is further separated by $\mathcal{S}^r$ and the component for the $j$-th speaker is $\mathbf{F}^r_{i,j}$:
\begin{equation}
    \begin{split}
    & \mathbf{M}_{i,1}^r, \dots, \mathbf{M}_{i,D}^r = \mathcal{S}^r (\mathbf{F}^r_i),\\
    & \mathbf{F}^r_{i,j} = \mathbf{F}^r_i \odot \mathbf{M}_{i,j}^r\ \ j = 1, \dots, D.
    \end{split}
\end{equation}

Accordingly, the refined encoded feature for $j$-th speaker is obtained by the summation over all $j$-th components (separated ingredients) from all speakers:
\begin{equation}
\label{eqn:refine_separator}
    \mathbf{G}^r_j = \sum_{i=1}^D \mathbf{F}^r_{i,j}.
\end{equation}

The decoder $\mathcal{D}^r$ in the refining phase is then employed to decode the refined features into speech sources $\hat{\mathbf{s}}^r_j \in \mathbb{R}^{1\times T}, j= 1, \dots, D$:
\begin{equation}
    \hat{\mathbf{s}}^r_j = \mathcal{D}^r (\mathbf{G}^r_j), j=1, \dots, D.
\end{equation}
The obtained speech sources $\hat{\mathbf{s}}^r$ are expected to be cleaner and more precise than the coarse version derived in the coarse phase due to the refining separation, which is verified in experiments of Section~\ref{sssec:ablation}.

\vspace{-3mm}
\subsection{Fine-grained Encoding Mechanism}
\label{ssec:FGEM}


The Fine-grained Encoding Mechanism is devised to construct a more fine-grained encoding space from the coarse encoding space for more precise and thorough separation among different speakers.
Specifically, we learn a new latent domain along each basis function of the existing latent domain that defines the coarse embedding space. Consequently, the derived new latent domain and the old latent domain jointly form a high-order domain, which defines a fine-grained encoding space by decomposing the coarse embedding space along the newly learned latent domain.


The coarse encoder $\mathcal{E}^c$ in the coarse phase transforms non-linearly the input speech signal into an encoding space by learning $N^c$ 1-D convolutional filters \maRevise{together with \emph{ReLU}}. These $N^c$ convolutional filters can be viewed as basis functions to form a latent domain denoted as $\mathcal{H}^c$. As a result, the learned encoding space in the coarse phase can be considered to be defined by such latent domain $\mathcal{H}^c$ comprising $N^c$ basis functions. We adopt the similar way to construct a new latent domain $\mathcal{H}^r$ upon $\mathcal{H}^c$. Formally, we learn a 1-D convolutional layer composed of $N^r$ filters of kernel size $1\times K^r$ together with a nonlinear activation layer \emph{ReLU} to construct a new latent domain with $N^r$ basis functions. The newly constructed latent domain $\mathcal{H}^r$ is applied to \textbf{each of $N^c$ basis functions} of the old domain $\mathcal{H}^c$ for a fine-grained decomposition. Consequently, the new domain $\mathcal{H}^r$ enables one more order of partitions of the encoding space upon the old domain $\mathcal{H}^c$, and thus a high-order (2-order) latent domain $\mathcal{H}_{\text{high}}$ is constructed. The $N^c$ basis functions of $\mathcal{H}^c$ and the $N^r$ basis functions of $\mathcal{H}^r$ correspond to basis functions in different orders of $\mathcal{H}_{\text{high}}$ respectively: 
one may view $\mathcal{H}_{\text{high}}$ as characterizing the encoding space by $N^c \times N^r$ individual features to achieve a fine-grained encoding space. 

\smallskip\noindent\textbf{Mathematical formulation.} The whole procedure is mathematically formulated as: 
\begin{equation}
\resizebox{1.0\hsize}{!}{$
    \begin{split}
    &N^c \text{ basis functions of domain $\mathcal{H}^c$ } \coloneqq N^c \text{ conv-filters of } \mathcal{E}^c,\\
    &N^r \text{ basis functions of domain $\mathcal{H}^r$ } \coloneqq N^r \text{ conv-filters of } \mathcal{E}^r,\\
    &\text{high-order domain $\mathcal{H}_{\text{high}}$} \coloneqq \mathcal{H}^r \times \mathcal{H}^c,\\
    &\mathbf{F}^r_{i}(n) = \mathcal{E}^r (\mathbf{F}^c_{i} (n)), i = 1, \dots, D; \ \ n = 1, \dots, N^c.\\
    &\mathbf{Feat}_{n, m} = \mathbf{F}^r_{i}(n, m), i = 1, \dots, D; \ \ n = 1, \dots, N^c; \ \ m = 1, \dots, N^r.
    \end{split}
    $}
\end{equation}
Herein, $\mathcal{H}^r \times \mathcal{H}^c$ denotes that the domain $\mathcal{H}^r$ is applied to each basis function of $\mathcal{H}^c$ in parallel. $\mathbf{F}^c_{i}(n) \in \mathbb{R}^{1 \times T'}$ is the encoded representations along the $n$-th basis function for the $i$-th speaker in domain $\mathcal{H}^c$ in the coarse phase, while $\mathbf{Feat}_{n,m} \in \mathbb{R}^{1 \times T''}$ is the corresponding encoded representations in fine-grained embedding space along the $n$-th basis function in the first order and the $m$-th basis function in the second order of the high-order domain $\mathcal{H}_{\text{high}}$ (in the refining phase). 
It should be noted that all the basis feature functions in the old domain are decomposed by the same refining encoder $\mathcal{E}^r$ to ensure that all decompositions are operated in the same new latent domain. To reduce the model complexity and avoid potential overfitting, in practice we apply the similar idea as group convolution~\cite{group_wise_conv} to perform transformation in group-wise manner. Concretely, we first divide $N^c$ basis functions of the old domain into $P$ groups and then apply the same refining encoder $\mathcal{E}^r$ to each group $\mathbf{F}^c_{i} (p) \in \mathbb{R}^{\frac{N^c}{P} \times T'}$ to obtain a set of $N^r$ decomposed features $\mathbf{F}^r_{i} (p) \in \mathbb{R}^{N^r \times T''}$:
\begin{equation}
    \mathbf{F}^r_{i}(p) = \mathcal{E}^r (\mathbf{F}^c_{i} (p)), i = 1, \dots, D; \ \ p = 1, \dots, P.
\end{equation}
Consequently, the size of the resulting encoded representations in the new domain $\mathbf{F}^r$ is $N^r \times P \times T''$.

\smallskip\noindent\textbf{Physical interpretation.} The rationale behind this design is that different basis functions correspond to different components in the latent domain, which is similar to different frequency bands in the frequency domain. Thus the speech signal for a speaker is characterized by the distribution of different components (along different basis functions) in a latent domain. Nevertheless, we argue that the capacity of feature representation in one latent domain is not sufficient to separate all speakers perfectly. For those mixtures of speech signal that cannot be thoroughly separated in one latent domain, our designed fine-grained embedding space encoded by a high-order latent domain enables more precise separation.

\smallskip\noindent\textbf{Comparison with the modeling mechanism of scaling up the number of basis functions in the old 1-order domain.} A straightforward way to expand the capacity of feature representation in the embedding space is to directly scale up the number of basis functions in the old domain $\mathcal{H}^c$. Whilst it seems plausible, it has two limitations compared to our proposed Fine-grained Encoding Mechanism: 1) This mechanism can only improve the feature representation in the same embedding space along the old latent domain which is prone to be saturated and overfitting. By contrast, our model exploits a larger embedding space by learning a new latent domain and applying it to each basis function of the old domain to form a high-order latent domain. Thus the feature representation power is increased quadratically instead of linearly. 2) This mechanism increases the parameter size of the encoder proportionally to the scaling factor. Nevertheless, the parameter size of the encoder in our model grows linearly to the scaling factor of the representation capacity, which benefits from the design that the same refining encoder $\mathcal{E}^r$ encoding the new latent domain is applied to all the basis functions of the old domain.
    
\smallskip\noindent\textbf{Refining separator $\mathcal{S}^r$ in the fine-grained embedding space.} As shown in Figure~\ref{fig:refining} (b), the refining separator $\mathcal{S}^r$ 
processes the encoded features in the fine-grained embedding space in parallel for all groups of basis functions of the old domain $\mathcal{H}^c$, i.e., the $P$ feature blocks $\mathbf{F}^r_i(p) \in \mathbb{R}^{N^r \times T''}, p=1,\dots,P$ for the $i$-th speaker, are processed in parallel:
\begin{equation}
    \begin{split}
    & \mathbf{M}_{i,j}^r(p) = \mathcal{S}^r (\mathbf{F}^r_i(p)), \\
    & \mathbf{F}^r_{i,j}(p) = \mathbf{F}^r_i(p) \odot \mathbf{M}_{i,j}^r(p), \ \ p = 1, \dots, P.
    \end{split}
\end{equation}
The refining separator $\mathcal{S}^r$ shares the same model structure as the coarse separator $\mathcal{S}^c$: consisting of R stacked blocks of DPRNN~\cite{DPRNN} or DPTNET~\cite{DPTNet}. 
The refined separated representation for $j$-th speaker is obtained by the summation over all $j$-th components, namely the separated ingredient of $j$-th speaker, from all speakers in parallel along each of $P$ feature blocks. Thus, Equation~\ref{eqn:refine_separator} is implemented as: 
\begin{equation}
    \mathbf{G}_j^r(p) = \sum_{i=1}^D\mathbf{F}^r_{i,j}(p), \ \ p = 1, \dots, P.
\end{equation}

\noindent\textbf{Refining decoder $\mathcal{D}^r$ in the fine-grained embedding space.}
Since our fine-grained embedding space is constructed based upon the coarse embedding space, our refining decoder $\mathcal{D}^r$ performs decoding reversely. Hence, the refining decoder $\mathcal{D}^r$ is a two-stage decoder consisting of $\mathcal{D}^r_1$ and $\mathcal{D}^r_2$. Specifically, the sub-decoder $\mathcal{D}^r_1$ decodes the feature representation from fine-grained embedding space back to the coarse embedding space, which is applied to all feature groups in parallel. Then the sub-decoder $\mathcal{D}^r_2$ decodes features from the coarse embedding space to final speech signal for each speaker. The whole decoding procedure is carried out as follows: 
\begin{equation}
    \begin{split}
        & \mathbf{G}_j^c(p) = \mathcal{D}^r_1(\mathbf{G}_j^r(p)),\ \ p = 1, \dots, P, \\
        & \hat{\mathbf{s}}_j^r = \mathcal{D}^r_2(\mathbf{G}_j^c)
    \end{split}
\end{equation}
As indicated in Figure~\ref{fig:refining} (c), $\mathcal{D}^r_1$ is modeled by a 1-D transposed convolutional layer followed by a nonlinear activation layer \emph{ReLU}, and $\mathcal{D}^r_2$ is modeled by a 1-D transposed convolutional layer. 

\subsection{End-to-end Parameter Learning by Joint Supervision}
We optimize the whole model of \emph{SRSSN} in an end-to-end manner by performing supervised learning on both the coarse phase and the refining phase. Specifically, we employ the  scale-invariant source-to-noise ratio (SI-SNR)~\cite{SISNR} as the loss function, which is widely used and validated effectively for end-to-end speech separation~\cite{Conv-tasnet, DPRNN, MULCAT, DPTNet, SepFormer}. Formally, given a target signal (groundtruth) $\mathbf{s}$ and the estimated signal $\hat{\mathbf{s}}$, the SI-SNR is defined as follows:
\begin{equation}
    \begin{split}
    & \mathbf{s}_{\text{target}} = \frac{\langle \hat{\mathbf{s}}, \mathbf{s}\rangle}{{\Vert \mathbf{s}\Vert}^2} \mathbf{s},\\
    & \mathbf{s}_{\text{noise}} = \hat{\mathbf{s}} - \mathbf{s}_{\text{target}}, \\
    & \text{SI-SNR} (\hat{\mathbf{s}}, \mathbf{s}) \coloneqq 10 \log_{10} \frac{{\Vert \mathbf{s}_{\text{target}}\Vert}^2}{{\Vert \mathbf{s}_{\text{noise}}\Vert}^2},
    \end{split}
\end{equation}
where $\mathbf{s}_{\text{target}}$ is the projected correct ingredient of estimated signal $\hat{\mathbf{s}}$ on the target signal and $\mathbf{s}_{\text{noise}}$ is the residual noise ingredient of $\hat{\mathbf{s}}$.
Following~\cite{DPRNN, DPTNet}, we adopt utterance-level PIT~\cite{UPIT} to align multiple estimates for different target speakers in both the coarse phase and the refining phase. Consequently, our \emph{SRSSN} is optimized by maximizing the SI-SNR scores in both the coarse phase and the refining phase:
\begin{equation}
    \begin{split}
    & \mathcal{L}^c = -\max_{\pi^c \in \mathcal{P}} \frac{1}{D} \sum_{i=1}^D \text{SI-SNR}(\mathbf{s}_i, \hat{\mathbf{s}}_{\pi^c(i)}^{c}), \\
    & \mathcal{L}^r = -\max_{\pi^r \in \mathcal{P}} \frac{1}{D} \sum_{i=1}^D \text{SI-SNR}(\mathbf{s}_i, \hat{\mathbf{s}}_{\pi^r(i)}^{r}),\\
    & \mathcal{L} = \mathcal{L}^c + \mathcal{L}^r. \\
    \end{split}
\end{equation}
where $\mathcal{L}^c$ and $\mathcal{L}^r$ are the loss functions used in the coarse phase and refining phase, respectively. $\pi^c$ and $\pi^r$ are permutations from the set $\mathcal{P}$ of all $D!$ possible permutations among $D$ speakers.

\label{sec:methods}

\section{Experiments}
To evaluate the effectiveness of our proposed \emph{SRSSN}, we conduct three sets of experiments: 1) speech separation in the clean (noise-free) setting \maRevise{involving 2 and 3 speakers respectively}, 2) speech separation between 2 speakers in noisy and reverberant settings, and 3) speech recognition on separated speech signals decoded by methods for speech separation to evaluate the performance of speech separation indirectly. We also perform ablation study on the task of speech separation in the clean setting to investigate the effect of each proposed functional technique in our proposed \emph{SRSSN}. 
In each set of experiments, we evaluate the performance of our \emph{SRSSN} adopting the separator structure of DPRNN-TASNET~\cite{DPRNN} and DPTNET-TASNET~\cite{DPTNet} respectively to evaluate the robustness of our model utilizing different classical separator structures. These two versions of our model are denoted as DPRNN-\emph{SRSSN} and DPTNET-\emph{SRSSN} respectively.

\subsection{Experimental Setup}
\noindent\textbf{Evaluation Metrics.} 
For speech separation, we employ two standard metrics for evaluation, namely scale-invariant signal-to-noise ratio improvement ($\Delta$SI-SNR)~\cite{SISNR} and signal-to-distortion ratio improvement ($\Delta$SDR)~\cite{SDR}. 
Higher value of $\Delta$SI-SNR or $\Delta$SDR indicates higher quality of the separated results. For speech recognition, we employ Word Error Rate (WER) of the predicted transcripts relative to the reference transcripts for evaluation. Lower value of WER implies better recognition result and higher quality of the separated speech. 

\noindent\textbf{Implementation Details}
Our model is implemented in Pytorch framework~\cite{pytorch}. It is trained using Adam~\cite{adam} optimizer with a learning rate of $10^{-3}$ and a weight decay of $10^{-5}$ on 2-second temporal segments for 200 epochs. We clip all gradients to lie in the interval $[-5, 5]$ to avoid potential gradient explosion.
For the encoder $\mathcal{E}^c$ in the coarse phase, the number of filters $N^c$, kernel size $K^c$, and stride size are set to be 256, 16, and 8, respectively.
For the encoder $\mathcal{E}^r$ in the refining phase, the number of filters $N^r$, kernel size $K^r$, stride size, and number of groups $P$ are tuned to be 256, 2, 1, and 4, respectively. For both separators $\mathcal{S}^c$ and $\mathcal{S}^r$, the numbers of core blocks $R$ are set to 6 except for the experiment on ablation study (Section~\ref{sssec:ablation}), and the length of chunks $L$ is set to 100.
In the DPRNN blocks, each Bi-LSTM layer is equipped with 128 hidden units in each direction. 
In the DPTNET blocks, each improved Transformer layer~\cite{DPTNet} consists of a 4-head self-attention layer with total embedding dimension of 64, a Bi-LSTM layer with 128 hidden units in each direction, and a linear layer with 64 hidden units.


\subsection{Speech Separation in Clean Setting}
\label{ssec:sep_clean}
We first conduct experiments on speech separation in clean setting, i.e., no noise is contained in the mixture of speech signals except the signals of involved speakers to be separated. We first perform ablation study to investigate the effectiveness of the coarse-to-fine separation framework and the proposed Fine-grained Encoding Mechanism. Then we compare our \emph{SRSSN} to the state-of-the-art methods for speech separation in this experimental setting.

\noindent\textbf{Dataset.} 
We perform experiments on WSJ0-2mix and \maRevise{ WSJ0-3mix~\cite{DPCL}} in the clean setting, which is the reference mixed speech datasets for single-channel speech separation. WSJ0-2mix is generated from Wall Street Journal (WSJ) dataset~\cite{wsj0}, and consists of mixed speech utterances from two different speakers with random signal-to-noise ratio between 0 dB and 5dB. The data in WSJ0-2mix is split into three sets with duration of 30 hours, 10 hours and 5 hours for training, validation and test, respectively. 
\maRevise{WSJ0-3mix, containing three-speaker mixtures, is generated in a similar way as in WSJ0-2mix.}
We use the \textit{min} version of speech data with sampling rate of 8kHz, the benchmark for speaker separation, in which the longer utterance is trimmed to align the shorter utterance.

\begin{figure*}
  \centering
  \begin{minipage}[b]{0.48\linewidth}
    \centering
    \includegraphics[width=\textwidth]{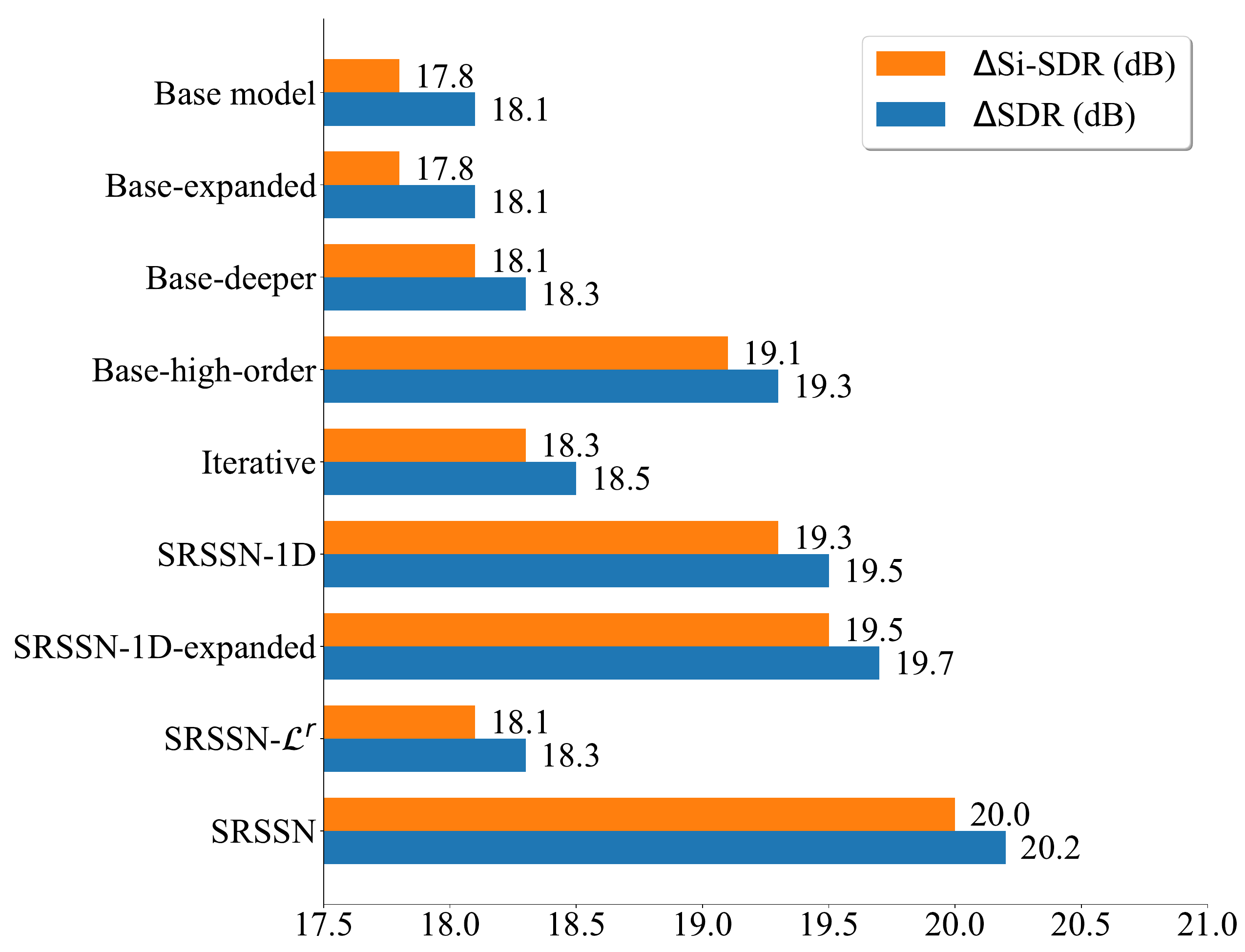}
    \centerline{(a) Performance of DPRNN-\emph{SRSSN}}\medskip
  \end{minipage}
  \hspace{0.01\linewidth}
  \begin{minipage}[b]{0.48\linewidth}
    \centering
    \includegraphics[width=\textwidth]{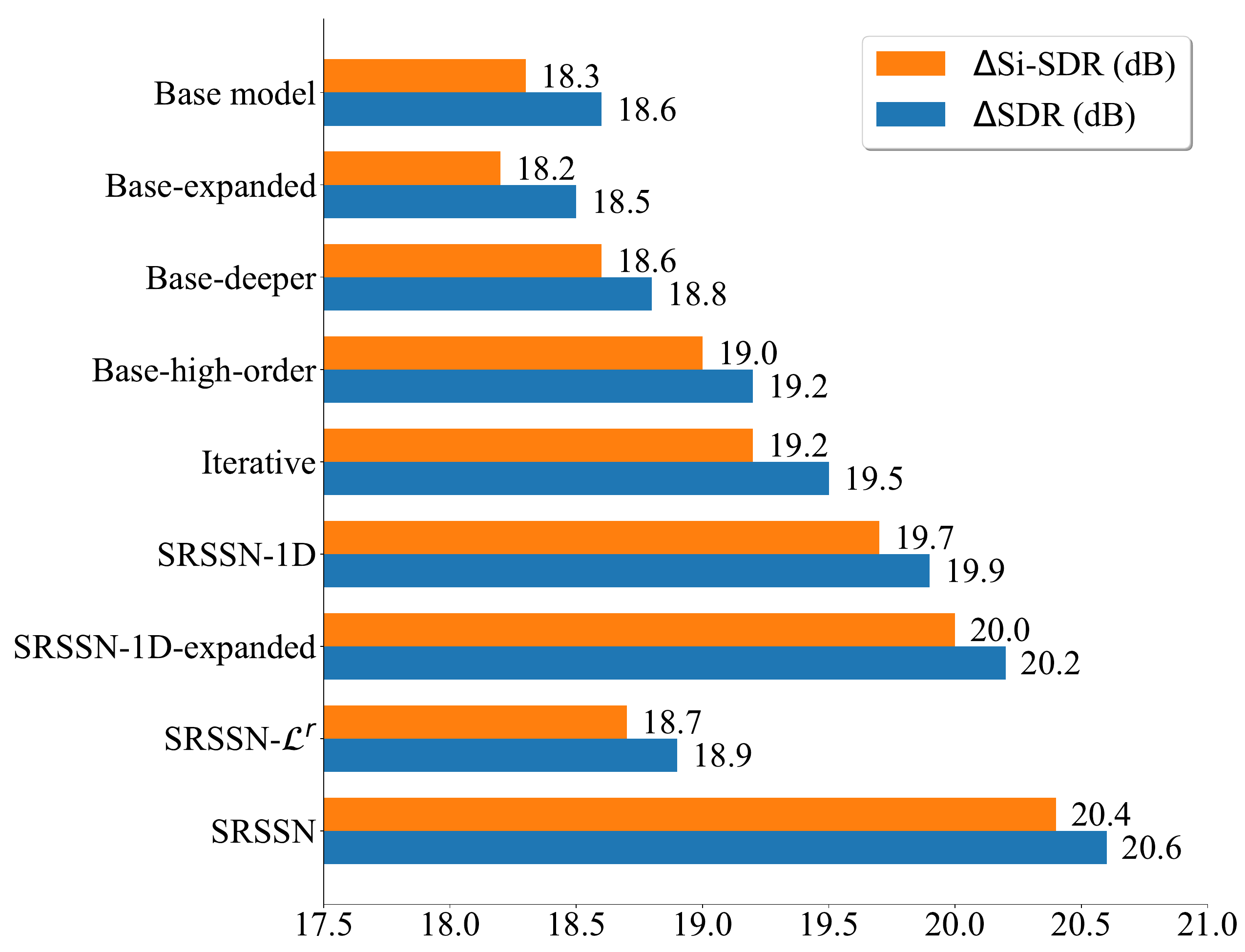}
    \centerline{(b) Performance of DPTNET-\emph{SRSSN}}\medskip
  \end{minipage}
  \caption{\maRevise{Performance of nine variants of our \emph{SRSSN} in terms of $\Delta$SI-SNR and $\Delta$SDR for ablation study, using DPRNN and DPTNET as separator respectively.}}
  \label{fig:ablation_groups}
\end{figure*}

\subsubsection{Ablation Study}
\label{sssec:ablation}
We perform ablation experiments on \maRevise{nine} variants of our \emph{SRSSN} for both DPRNN-\emph{SRSSN} and DPTNET-\emph{SRSSN} \maRevise{on WSJ0-2mix~\cite{DPCL}}:
\begin{itemize}\setlength{\itemsep}{-0cm}
    \item \textbf{Base model}, which has only coarse phase and thus no Fine-grained Encoding Mechanism is used. As a result, the base model is equivalent to DPRNN-TASNET or DPTNET-TASNET, depending on the separator structure (based on DPRNN or DPTNET blocks). The stride size for \textbf{Base model} in both cases are set to be 8, which is consistent with all other models in ablation study. 
    \item \textbf{Base-expanded}, which is similar to \textbf{Base model} with one difference: the encoding space in the coarse phase is expanded by scaling up the number of basis functions in the latent domain to exploit the limit of sufficiently large encoding space, namely using much more CNN filters ($N^c=1024$) for the encoder. 
    \maRevise{
    \item \textbf{Base-deeper}, which has deeper encoder and decoder than \textbf{Base model}, to investigate the effect of the convolutional depth of both the encoder and decoder. In this variant, we deepen the encoder and the decoder with 2, 3 and 4 convolutional layers (1-D) respectively, and select the best performance as the optimal results w.r.t. the convolutional depth.  
    }
    \maRevise{
    \item \textbf{Base-high-order}, which directly encodes the mixed speech into high-order embedding space and perform speech separation in only one separation phase. Compared to our \textbf{\emph{SRSSN}}, no rough separation is performed in the 1-order embedding space. Hence, this variant is proposed to validate the effectiveness of coarse-to-fine separation scheme. Specifically, the encoder is modeled by cascading the coarse encoder $\mathcal{E}^c$ and refining encoder $\mathcal{E}^r$, to encode the speech signal into the fine-grained embedding space. The refining separator $\mathcal{S}^r$ and the refining decoder $\mathcal{D}^r$ are applied subsequently. 
    }
    \item \textbf{Iterative}, which adopts the iterative scheme used in iTDCN++~\cite{universal, improving}. Specifically, the separation procedure in \textbf{Base model} is repeated twice, where the mixed speech and the initial estimates in the first phase are fed into the second phase. Two phases are cascaded into an end-to-end model.
    \item \textbf{\emph{SRSSN}-1D}, which has both the coarse and refining phases but no Fine-grained Encoding Mechanism is used in the refining phase. The refining phase has the same model structure as the coarse phase. Thus both two phases encode features in a 1-order latent domain. Note that \textbf{\emph{SRSSN}-1D} is different from \textbf{Iterative} in that the refining phase of \textbf{\emph{SRSSN}-1D} accepts the coarsely separated latent representations as input whilst the second phase of \textbf{Iterative} takes the decoded estimated signals as the input.
    \item \textbf{\emph{SRSSN}-1D-expanded}, which is similar to \textbf{\emph{SRSSN}-1D} with one difference: the encoding space in the refining phase is expanded by scaling up the number of basis functions in the latent domain. Similar to \textbf{Base-expanded}, much larger number of CNN filters ($N^r=1024$) are used for the refining encoder.
    \item \textbf{\emph{SRSSN}-$\mathcal{L}^r$}, which is the same as the proposed $\emph{SRSSN}$, except that the model is trained with only loss function $\mathcal{L}^r$ in the refining phase, discarding the loss function $\mathcal{L}^c$ in the coarse phase. 
    \item \textbf{\emph{SRSSN}}, which is our intact model: the Coarse-to-fine framework is applied and the Fine-grained Encoding Mechanism is leveraged to construct a fine-grained encoding space defined by a learned high-order latent domain, which enables fine-grained separation.
\end{itemize}
Since the separator accounts for most of model parameters, the total number of core blocks (DPRNN or DPTNET) in the separator for each variant is kept consistent for a fair comparison. Specifically, for the variants with only one separation phase including \textbf{Base model}, \textbf{Base-expanded}, \maRevise{\textbf{Base-deeper}, and \textbf{Base-high-order},} the number of core blocks $R$ is set to 6. For the variants with two separation phases, including \textbf{Iterative}, \textbf{\emph{SRSSN}-1D}, \textbf{\emph{SRSSN}-1D-expanded}, \textbf{\emph{SRSSN}-$\mathcal{L}^r$}, and \textbf{\emph{SRSSN}}, the number of blocks $R$ in both phases is set to 3. Figure~\ref{fig:ablation_groups} presents the experimental results of these \maRevise{nine} variants of our model with two different separator structures (DPRNN-based and DPTNET-based) for ablation study.

\maRevise{It should be noted that the convolutional sampling rate (on the input signal) in the refining phase of our model are actually equal to the coarse phase when setting the stride size in the refining phase to be 1, because the refining phase is performed on the output of the coarse phase. In all our implementation, the stride size in the coarse phase and refining phase are set to be 8 and 1 respectively, thus the overall stride size (on the input signal) of our \textbf{\emph{SRSSN}} after two phases is 8, which is equal to other methods in the ablation study. In such experimental settings, the comparisons between our model and other models in ablation study are fair.}

\begin{figure*}[t]
  \centering
  \begin{minipage}[b]{0.44\linewidth}
    \centering
    \includegraphics[width=\textwidth]{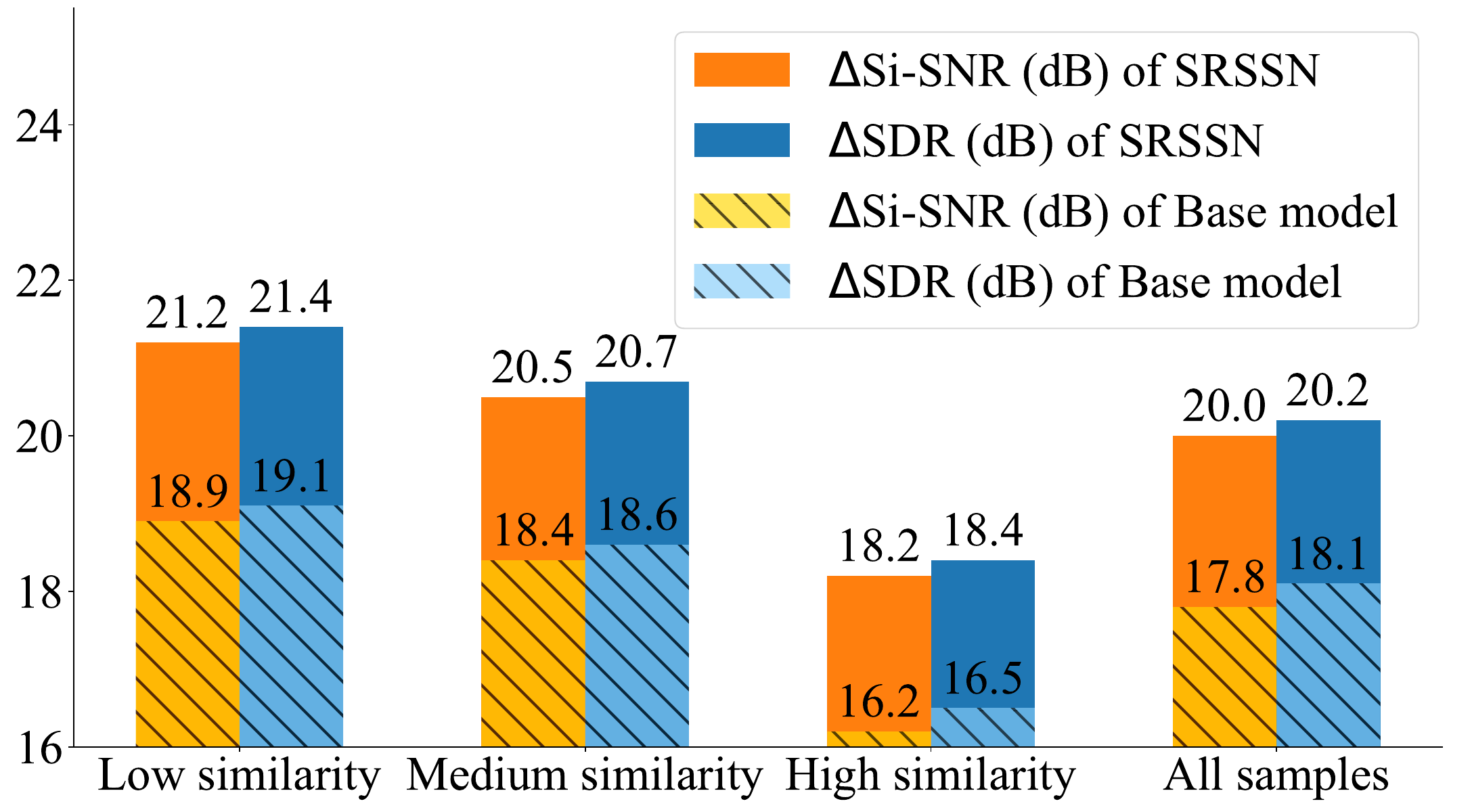}
    \centerline{(a) Performance of using DPRNN as separator}\medskip
  \end{minipage}
  \hspace{0.03\linewidth}
  \begin{minipage}[b]{0.44\linewidth}
    \centering
    \includegraphics[width=\textwidth]{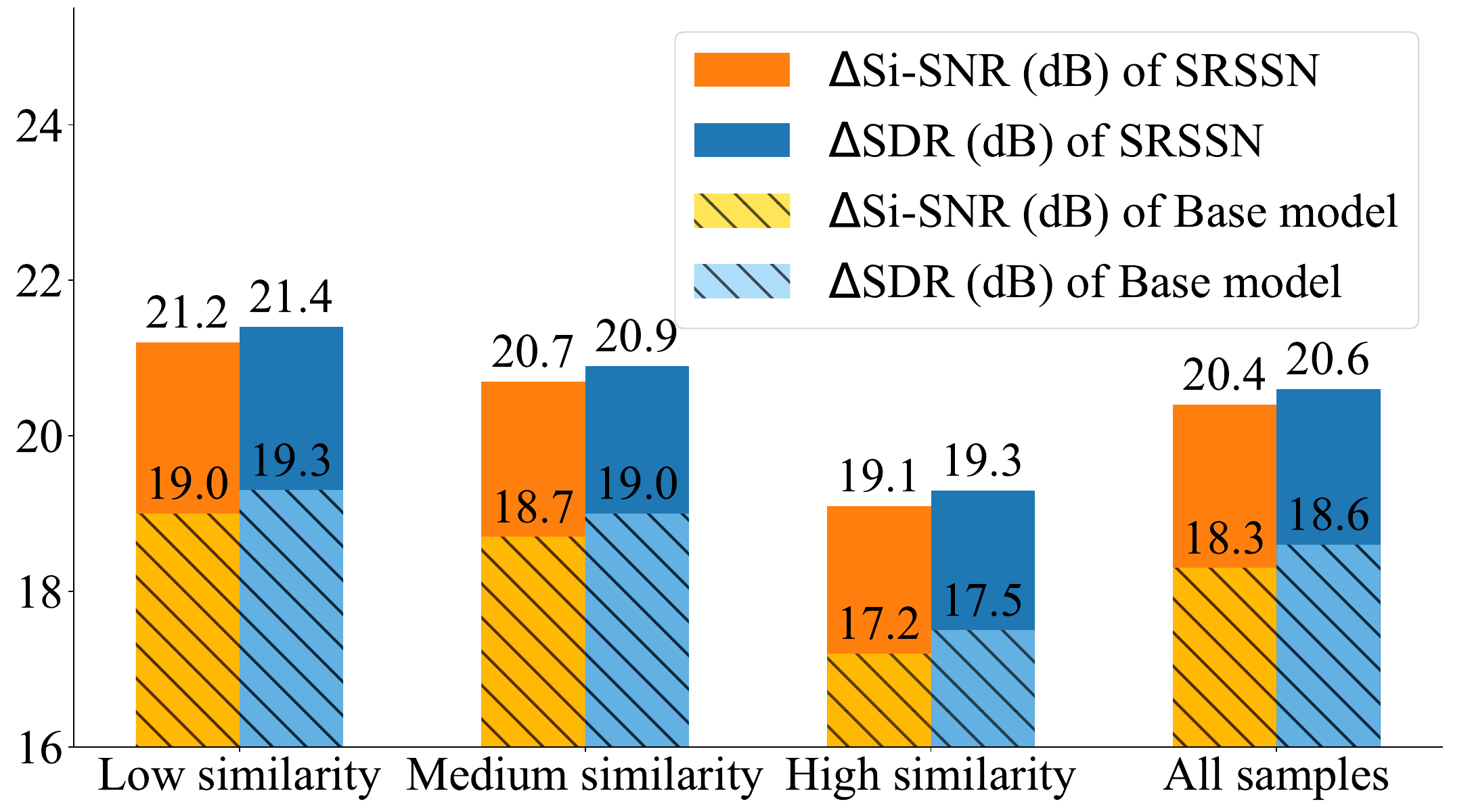}
    \centerline{(b) Performance of using DPTNET as separator}\medskip
  \end{minipage}
  \caption{\maRevise{Performance of our \textbf{\emph{SRSSN}} and \textbf{Base model} in terms of $\Delta$ SI-SNR and $\Delta$ SDR on different test subsets with different level of similarity between involved speakers in the mixed speech.}}
  \label{fig:diff_data}
\end{figure*}

\smallskip\noindent\textbf{Effect of Coarse-to-fine framework.}
For both DPRNN-based and DPTNET-based separator structures, the performance is improved significantly in both metrics from \textbf{Base model} to \textbf{\emph{SRSSN}-1D}, which manifests the remarkable advantages of our proposed Coarse-to-fine separation framework. Although the coarse phase and the refining phase of \textbf{\emph{SRSSN}-1D} have the same model structure especially with the same encoding scheme, the learned encoding features of two phases are able to adapt to different separation stages under the guidance of the loss functions during training. The performance comparison between \textbf{Iterative}~\cite{universal, improving} and \textbf{\emph{SRSSN}-1D} demonstrates the benefit of our proposed Coarse-to-fine framework. The strategy of progressive separation through multiple phases is also adopted in \textbf{Iterative}, where the initial estimates from the first phase serve as prior speaker information to improve the separation performance in the second phase. However, it needs to learn a new encoding space for separation from scratch. In our proposed Coarse-to-Fine method, the separated representations from the first phase are further separated in the second phase, where the more thorough encoding space is constructed based on the existing coarse encoding space.
The method \textbf{\emph{SRSSN}-$\mathcal{L}^r$} trained with only loss function in the refining phase obtained much lower performance than the final version. The reason is that the performance of the coarse phase degenerates notably without direct supervision by the loss function on it.

\maRevise{
The performance gain from \textbf{High-order} to \textbf{\emph{SRSSN}} also indicates the advantages of the progressive separation strategy through multiple phases defined in our proposed Coarse-to-fine framework.}


\smallskip\noindent\textbf{Effect of Fine-grained Encoding Mechanism.}
Our intact \textbf{\emph{SRSSN}} outperforms \textbf{\emph{SRSSN}-1D} substantially in both DPRNN-based and DPTNET-based cases, which indicates the effectiveness of the proposed Fine-grained Encoding Mechanism. Compared to the 1-order latent domain in \textbf{\emph{SRSSN}-1D}, the constructed high-order latent domain space in \textbf{\emph{SRSSN}} enables the model to perform separation in more fine-grained encoding space and achieve more precise separation results. 
To further explore the performance limit of 1-order latent domain and investigate the essential difference between 1-order and high-order latent domain space, we scale up the number of basis functions in the latent domain of \textbf{Base model} and \textbf{\emph{SRSSN}-1D} and compare our \textbf{\emph{SRSSN}} with the expanded models \textbf{Base-expanded} and \textbf{\emph{SRSSN}-1D-expanded}.  Figure~\ref{fig:ablation_groups} shows similar results in \textbf{Base-expanded} and \textbf{\emph{SRSSN}-1D-expanded}. Increasing the number of basis functions in 1-order latent space slightly improves the separation performance in the case of \textbf{\emph{SRSSN}-1D-expanded} at the cost of proportionally increasing the parameters of encoder. In the case of DPTNET-\emph{SRSSN}, the performance is even degenerated from \textbf{Base model} to \textbf{Base-expanded} due to potential overfitting in 1-order latent domain. Our model \textbf{\emph{SRSSN}} performs distinctly better than these expanded models. 

\maRevise{
As shown in Figure~\ref{fig:ablation_groups}, \textbf{Base-deeper}, which has deeper encoder and decoder than \textbf{Base model},  slightly improves the separation performance than \textbf{Base model}. However, it performs worse than \textbf{Base-high-order}, which reveals that the high-order encoding space modeled by our model is not equivalent to (but more powerful than) the deeper encoding space by \textbf{Base-deeper}. 
}
The high-order latent space constructed by the fine-grained encoding mechanism significantly improves the feature representation power with even less encoder parameters due to the constructed mechanism of high-order domain described in \ref{ssec:FGEM}.

\begin{figure*}[!t]
  \centering
  \begin{minipage}[b]{0.42\linewidth}
    \centering
    \includegraphics[width=\textwidth]{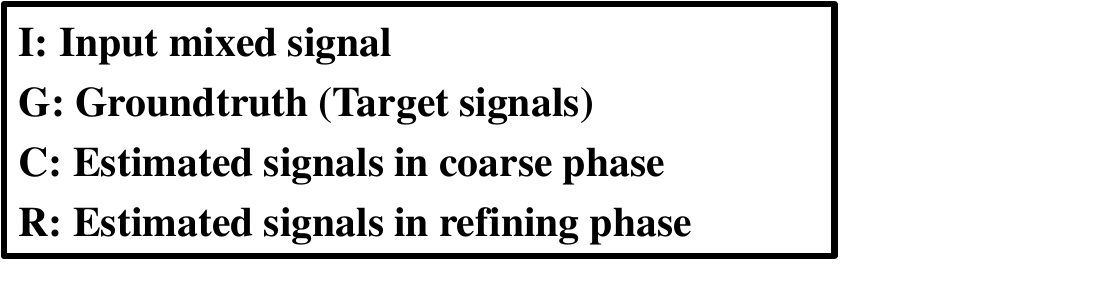}
  \end{minipage}
  \hspace{0.05\linewidth}
  \begin{minipage}[b]{0.42\linewidth}
    \centering
    \includegraphics[width=\textwidth]{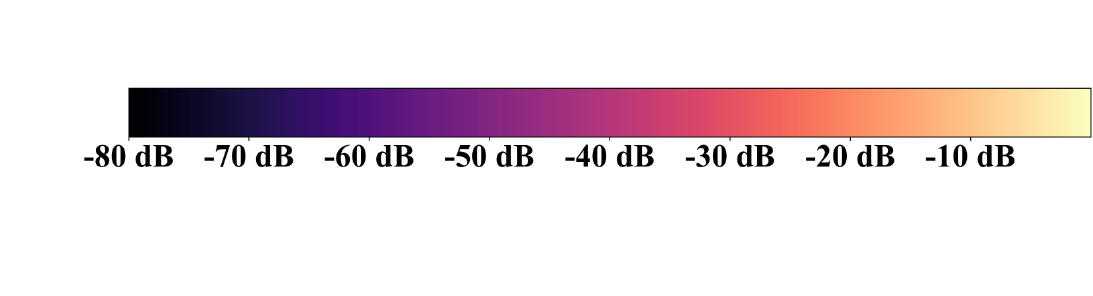}
  \end{minipage}
  \begin{minipage}[b]{0.42\linewidth}
    \centering
    \includegraphics[width=\textwidth]{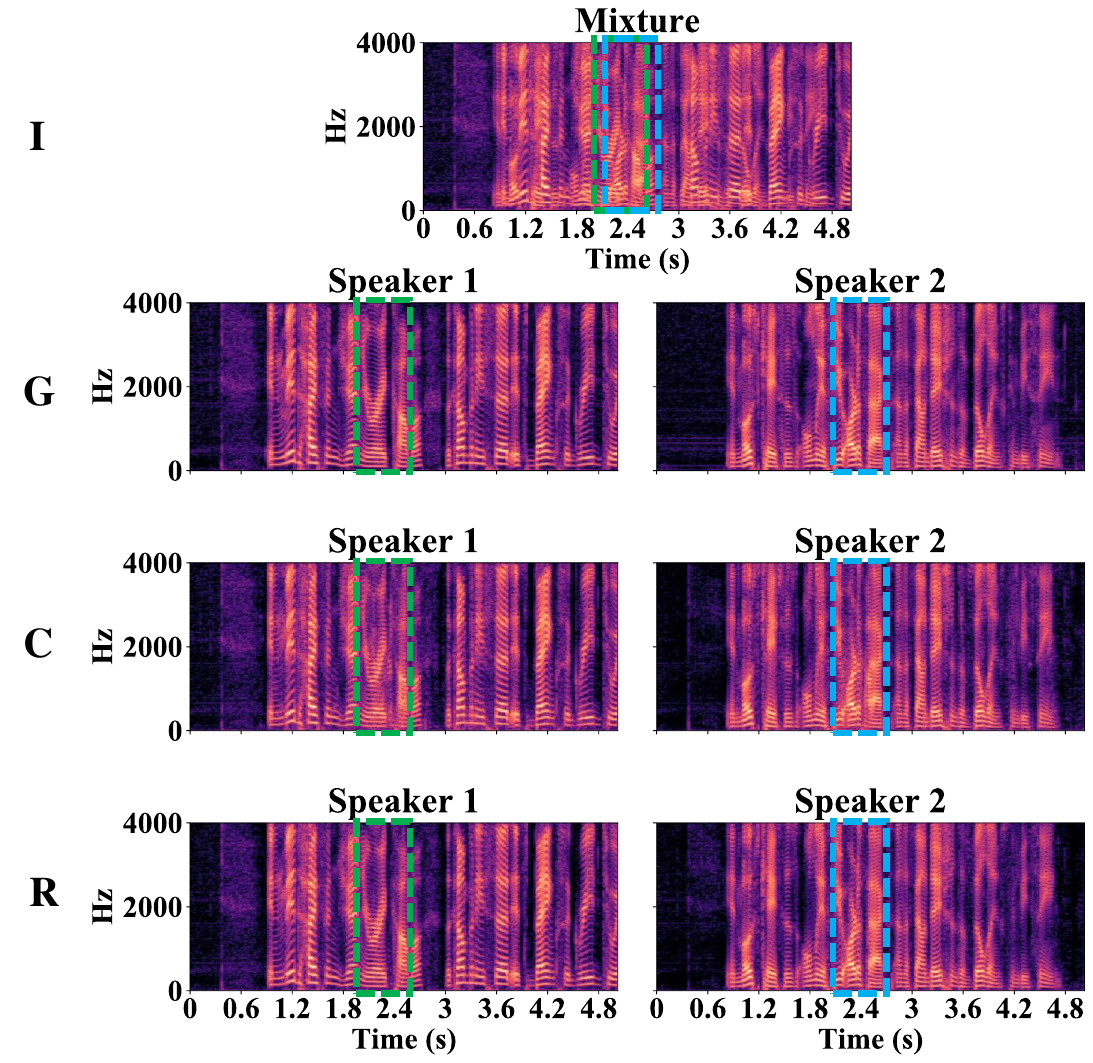}
    \centerline{(a) Sample 1 in DPRNN-\emph{SRSSN}}\medskip
  \end{minipage}
  \hspace{0.05\linewidth}
  \begin{minipage}[b]{0.42\linewidth}
    \centering
    \includegraphics[width=\textwidth]{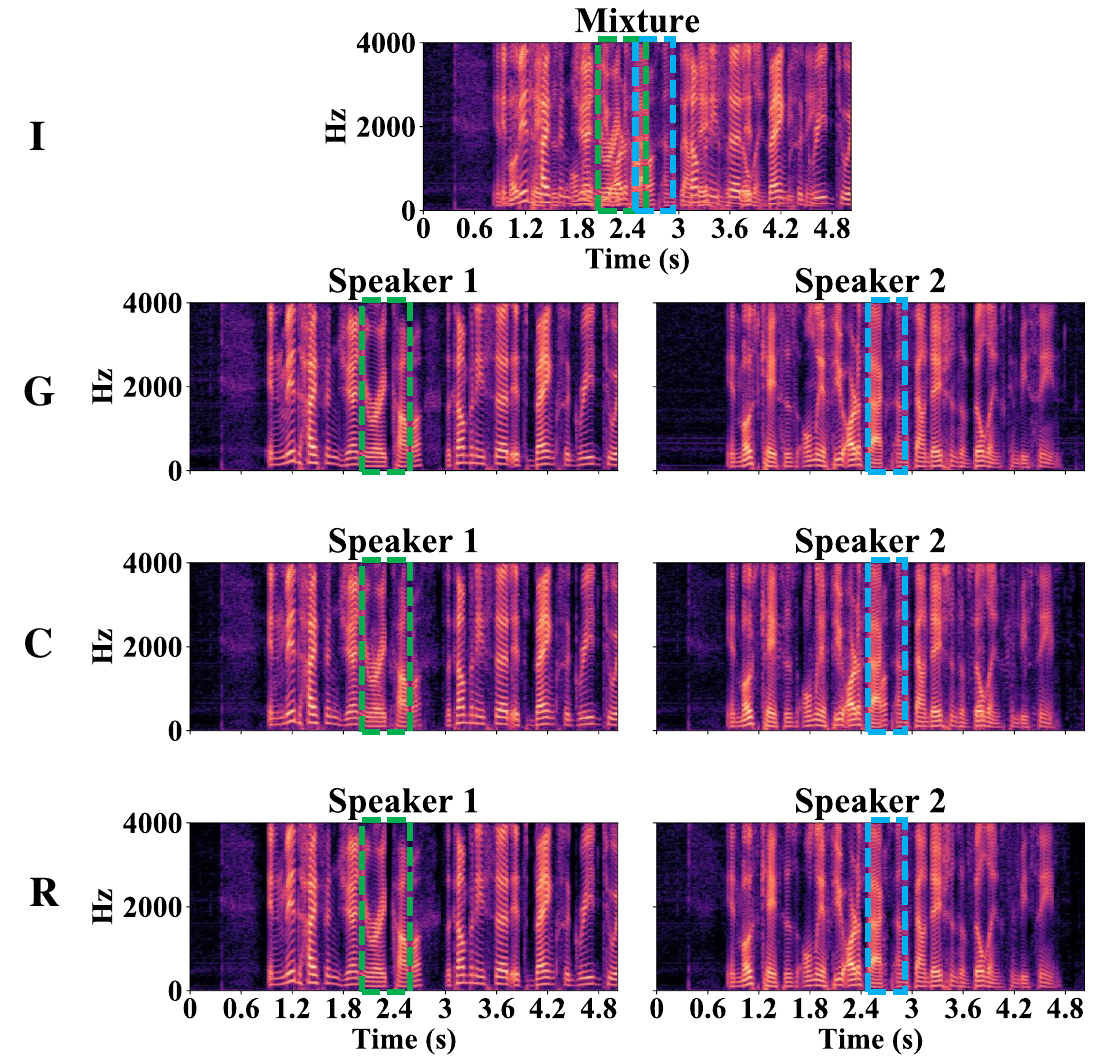}
    \centerline{(b) Sample 1 in DPTNET-\emph{SRSSN}}\medskip
  \end{minipage}
  \begin{minipage}[b]{0.42\linewidth}
    \centering
    \includegraphics[width=\textwidth]{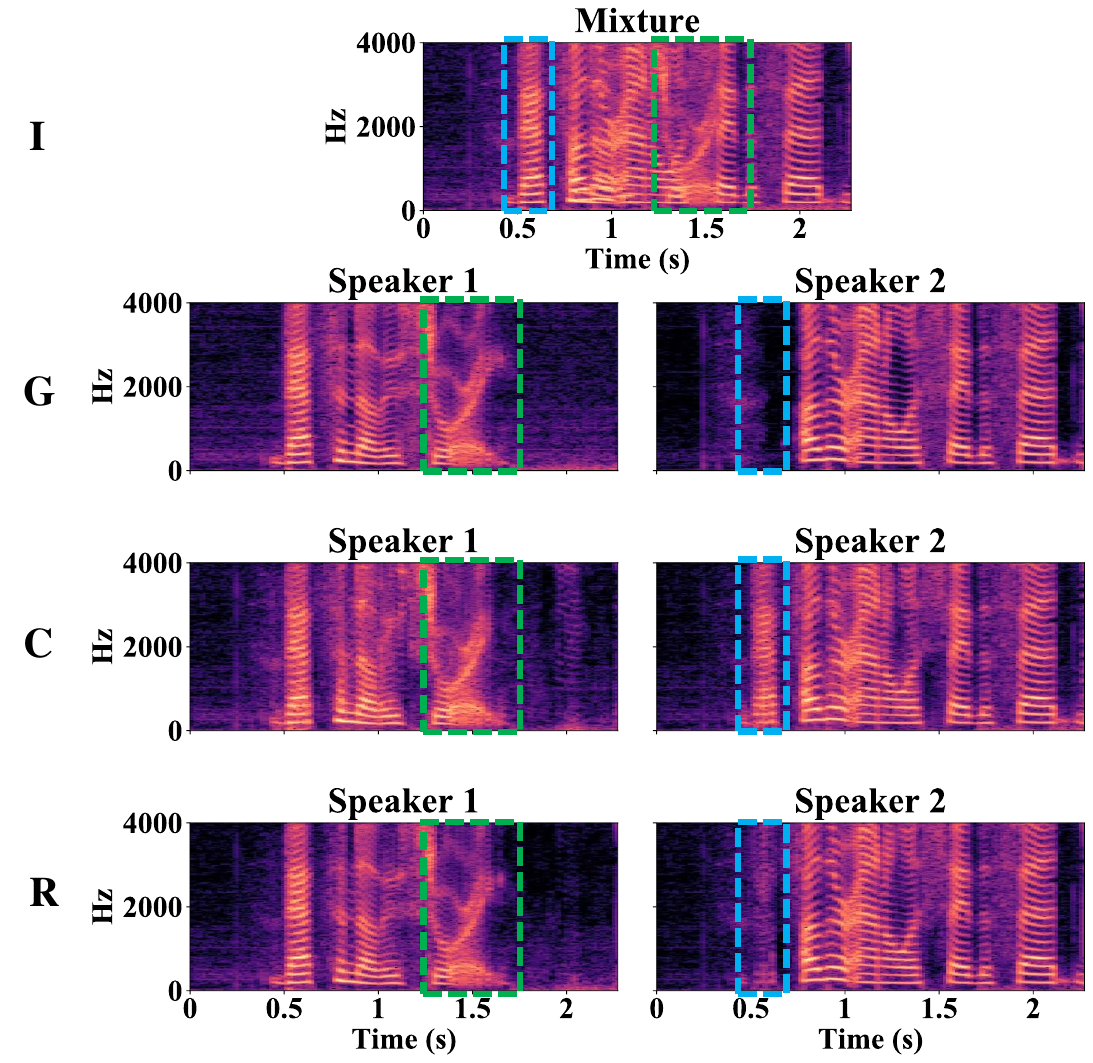}
    \centerline{(c) Sample 2 in DPRNN-\emph{SRSSN}}\medskip
  \end{minipage}
  \hspace{0.05\linewidth}
  \begin{minipage}[b]{0.42\linewidth}
    \centering
    \includegraphics[width=\textwidth]{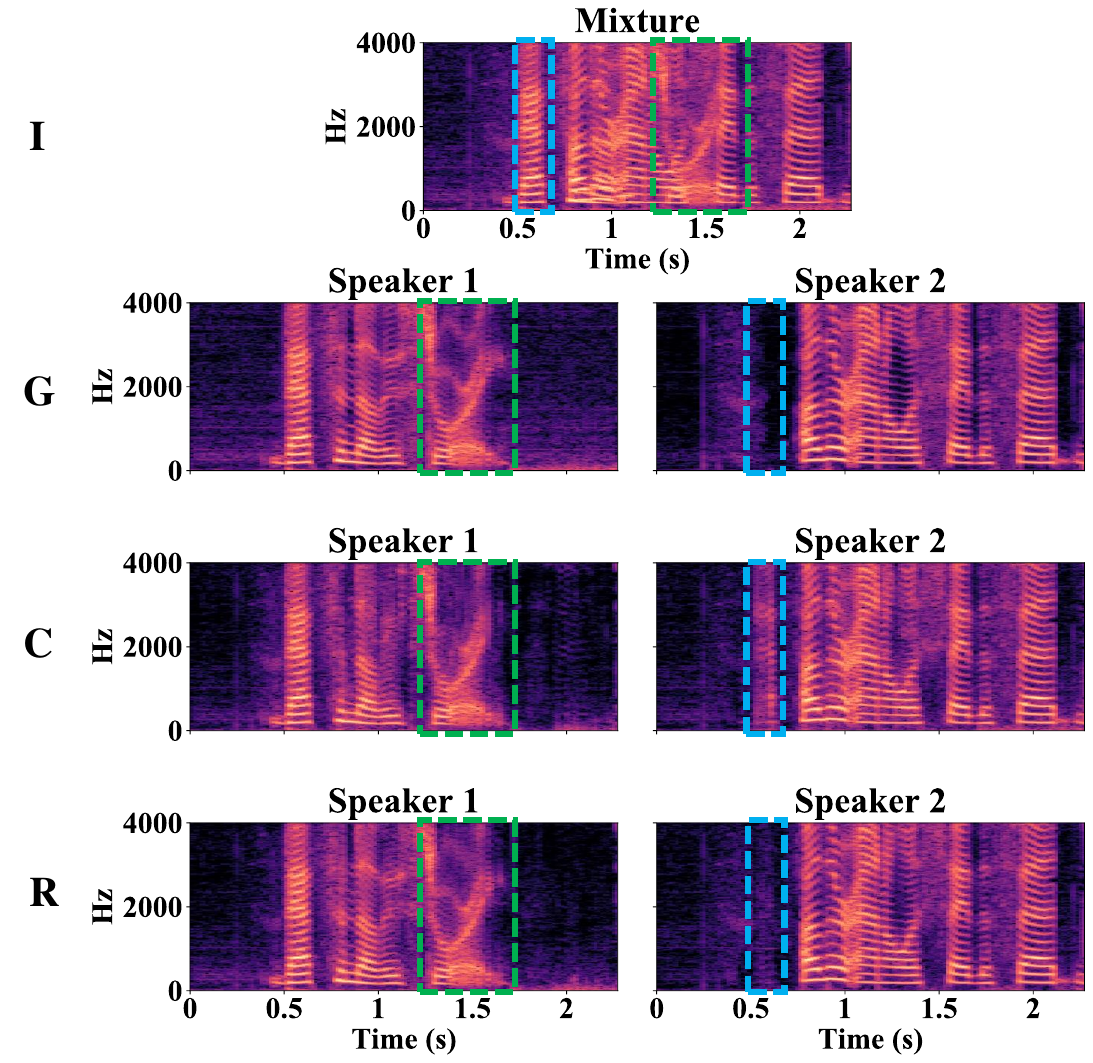}
    \centerline{(d) Sample 2 in DPTNET-\emph{SRSSN}}\medskip
  \end{minipage}
  \caption{Visualization of STFT power spectrums of our DPRNN-\emph{SRSSN} (left) and DPTNET-\emph{SRSSN} (right) in both coarse and refining phases on two randomly selected samples from test set. The contrasting regions are highlighted in green and blue boxes.}
  \label{fig:real_examples}
\end{figure*}

\begin{figure*}[!t]
  \centering
  \begin{minipage}[b]{0.42\linewidth}
    \centering
    \includegraphics[width=\textwidth]{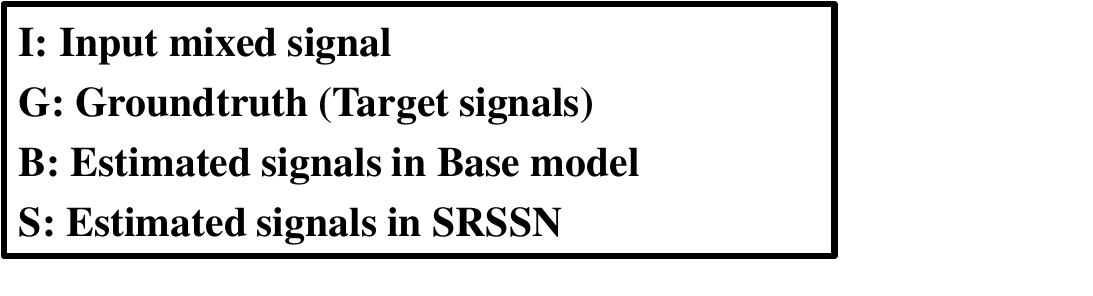}
  \end{minipage}
  \hspace{0.05\linewidth}
  \begin{minipage}[b]{0.42\linewidth}
    \centering
    \includegraphics[width=\textwidth]{spec_bar}
  \end{minipage}
  \begin{minipage}[b]{0.42\linewidth}
    \centering
    \includegraphics[width=\textwidth]{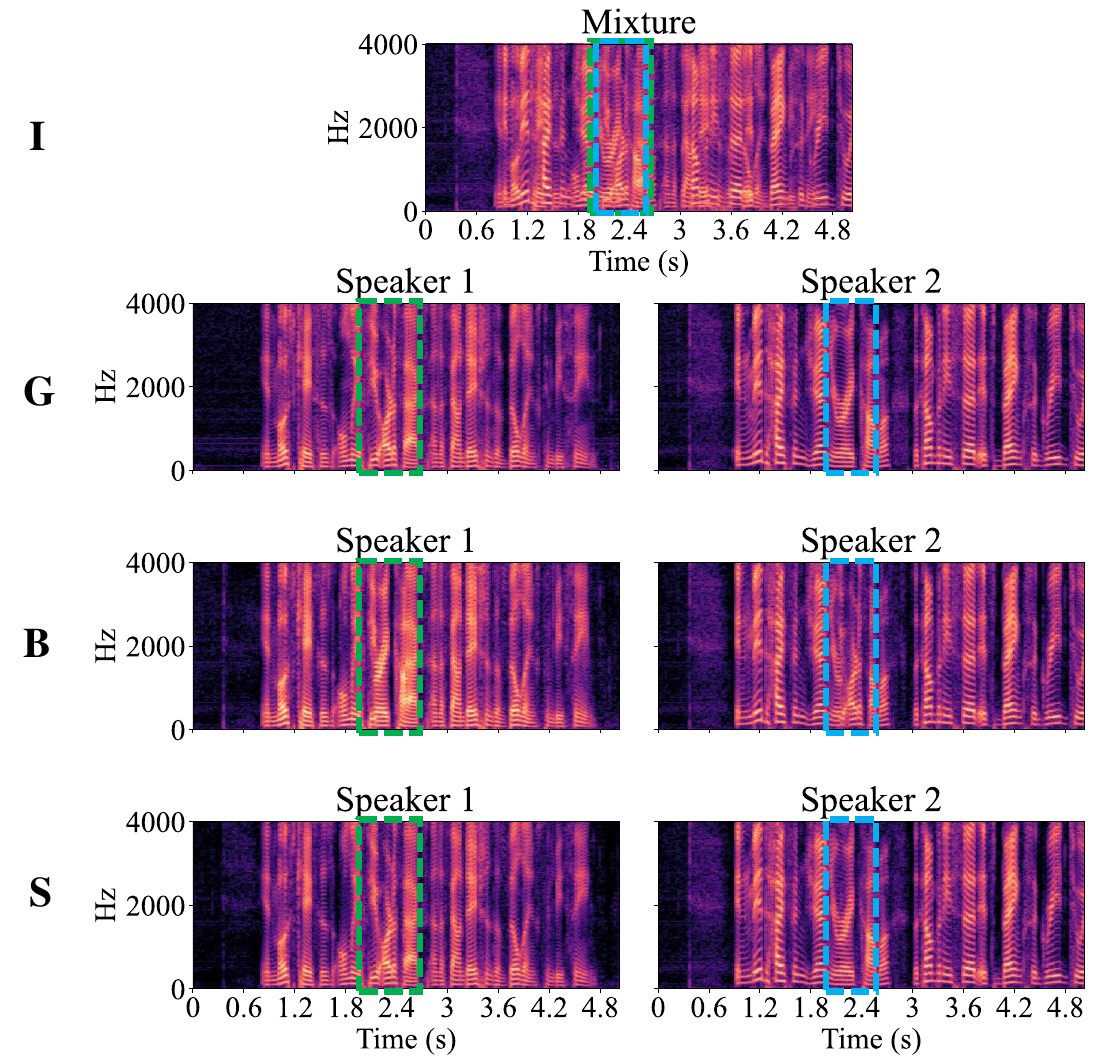}
    \centerline{(a) Sample 1 in case of DPRNN}\medskip
  \end{minipage}
  \hspace{0.05\linewidth}
  \begin{minipage}[b]{0.42\linewidth}
    \centering
    \includegraphics[width=\textwidth]{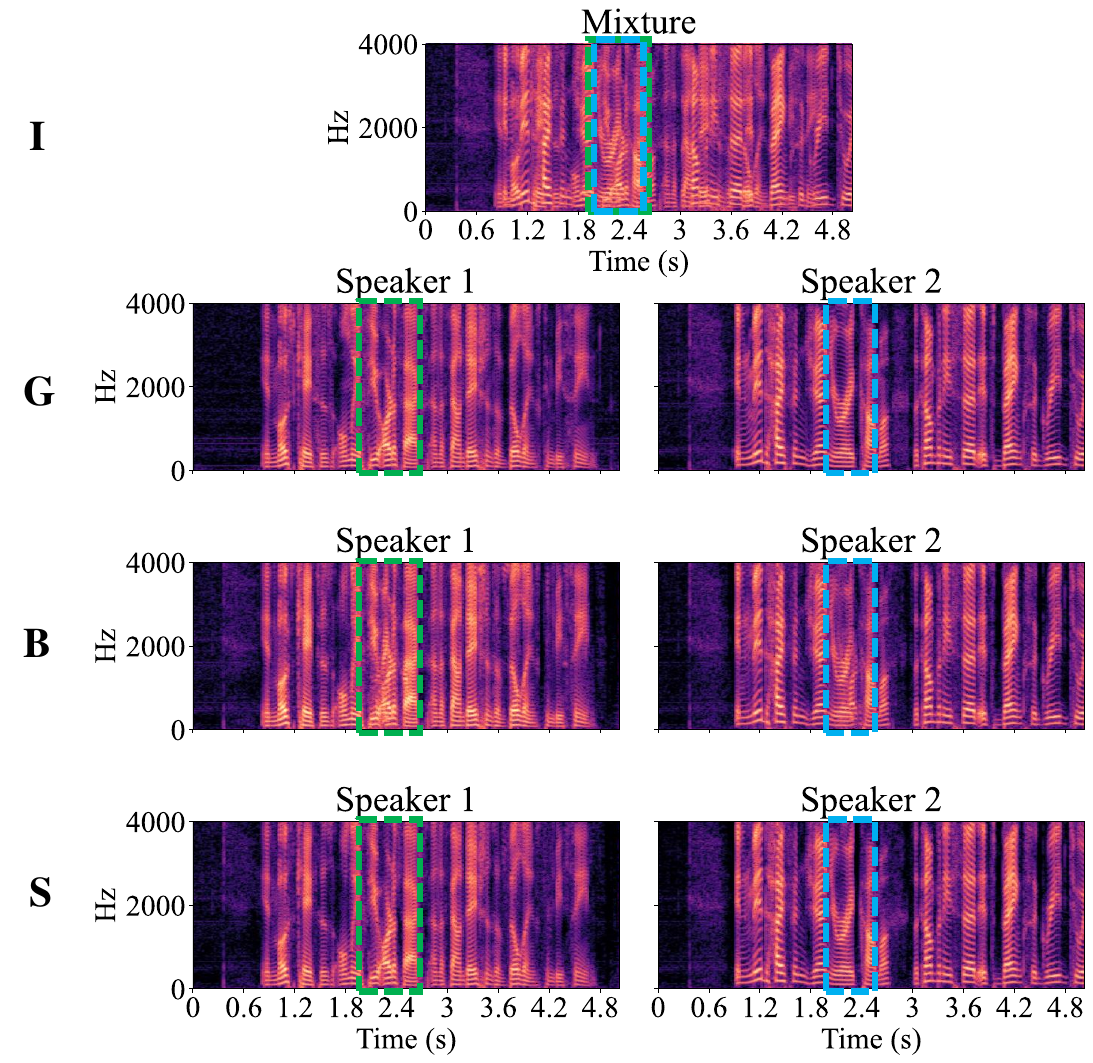}
    \centerline{(b) Sample 1 in case of DPTNET}\medskip
  \end{minipage}
  \begin{minipage}[b]{0.42\linewidth}
    \centering
    \includegraphics[width=\textwidth]{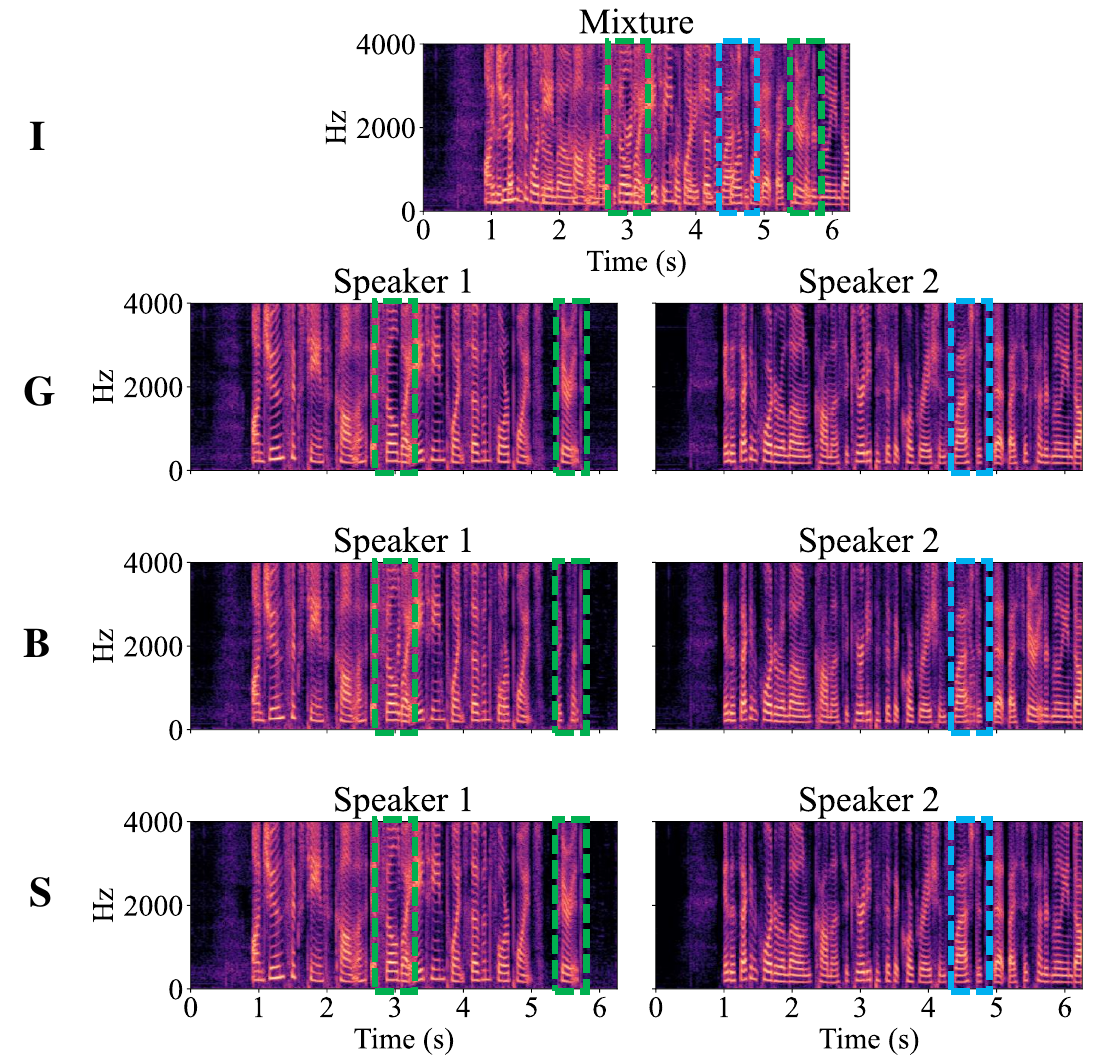}
    \centerline{(c) Sample 2 in case of DPRNN}\medskip
  \end{minipage}
  \hspace{0.05\linewidth}
  \begin{minipage}[b]{0.42\linewidth}
    \centering
    \includegraphics[width=\textwidth]{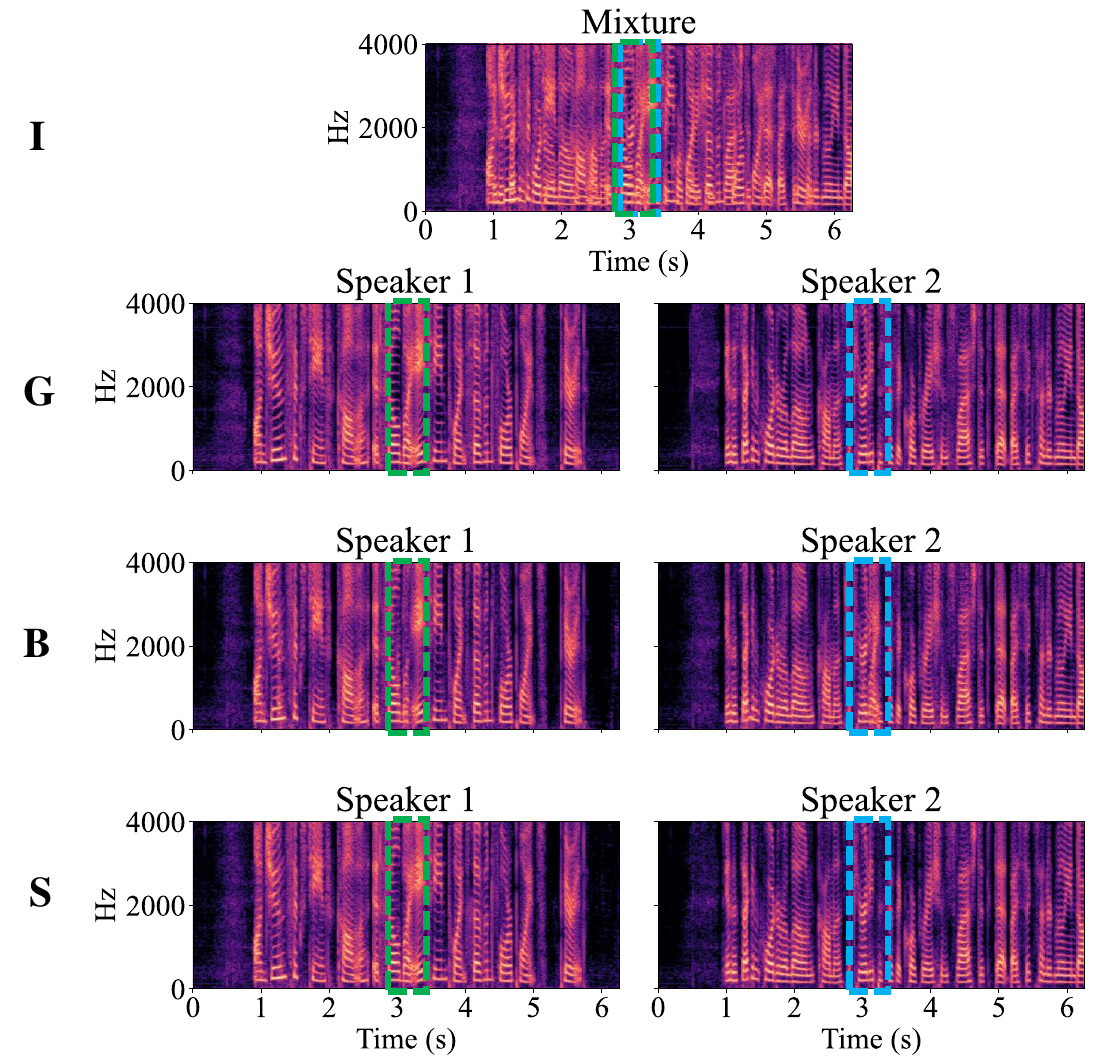}
    \centerline{(d) Sample 2 in case of DPTNET}\medskip
  \end{minipage}
  \caption{\maRevise{Visualization of STFT power spectrums of our \textbf{\emph{SRSSN}} and \textbf{Base model} in case of DPRNN and DPTNET on two randomly selected samples from the subset of \textit{High similarity} in test set. The contrasting regions are highlighted in green and blue boxes.}}
  \label{fig:real_examples_diff}
\end{figure*}

\smallskip\noindent\textbf{\maRevise{Robustness of the proposed \textbf{\emph{SRSSN}}}.}
\maRevise{To evaluate the robustness of our \textbf{\emph{SRSSN}}, we divide the test data into separate subsets according to the similarity between involved two speakers and compare the performance between our \textbf{\emph{SRSSN}} and \textbf{Base model}. Typically, mixed speech with more similarities between speakers is harder to be separated. 
Specifically, we utilize a pre-trained speaker encoder model\footnote[1]{\maRevise{https://github.com/resemble-ai/Resemblyzer}}
~\cite{speaker-enc} to extract the embedding vector of each involved speaker and calculate their cosine similarity for each sample. Higher cosine similarity indicates higher similarity of speech characteristics of involved speakers. We sort the samples in test set according to the similarity, and divide them into three subsets with equal number of samples, namely \textit{Low similarity}, \textit{Medium similarity} and \textit{High similarity}. Figure~\ref{fig:diff_data} presents the results on these three subsets and all samples. As the speaker similarity increases, the performance decreases for both models, which is reasonable. Our \textbf{\emph{SRSSN}} consistently outperforms \textbf{Base model} on all subsets in both metrics, which manifests the robustness of our \textbf{\emph{SRSSN}}.
}

\begin{table}[t]
  \centering
  \caption{\maRevise{Performance of different number of phases of stepwise separation in terms of $\Delta$SI-SNR (dB) and $\Delta$SDR (dB) using DPRNN as separator.}}
  \label{tab:comparison_stages}
  \renewcommand{\arraystretch}{1.4}
  \resizebox{0.99\linewidth}{!}{
  \begin{tabular}
  {lccccc}
  \toprule
  Method & Model size & GPU Memory usage & $\Delta$SI-SNR $\uparrow$ & $\Delta$SDR $\uparrow$ \\
  \midrule
  1-phase & 2.5M & 1.76Gib & 17.5 & 17.7 \\ 
  2-phase & 2.7M & 2.07Gib & 19.0 & 19.3 \\ 
  3-phase & 3.0M & 3.86Gib & 19.1 & 19.3 \\ 
  \bottomrule 
  \end{tabular} }
\end{table}

\smallskip\noindent\textbf{\maRevise{Exploration of number of phases for stepwise separation.}}
\maRevise{Theoretically, our proposed Fined-grained Encoding Mechanism can be iteratively employed without limitation to constructed higher-order encoding space and perform more fine-grained speech separation. However, more times of iterations inevitably lead to the increase of the model complexity and computational cost but diminishing marginal performance gain. We conduct experiments to investigate the scalability of our \textbf{\emph{SRSSN}} with increasing separation phases: 1) \textbf{1-phase}, which is the same as \textbf{Base model}; 2) \textbf{2-phase}, which is the same as current version of our \textbf{\emph{SRSSN}}; 3) \textbf{3-phase}, which performs stepwise separation through 3 phases sequentially in the 1-order, 2-order, and 3-order encoding space, respectively. Due to the memory limit, we only perform experiments in the case of DPRNN separator and use fewer DPRNN blocks. For \textbf{1-phase}, we use 4 blocks. For \textbf{2-phase}, we use 2 blocks for both phases. For \textbf{3-phase}, we use 2 blocks for the first phase and 1 phase for the later two phases. Similar to the construction of the 2-order embedding space, the 3-order embedding space is constructed by the decomposition of the 2-order embedding space using our proposed Fined-grained Encoding Mechanism.}

\maRevise{Table~\ref{tab:comparison_stages} presents the separation performance of these versions of \textbf{\emph{SRSSN}}. The performance is improved significantly from \textbf{1-phase} to \textbf{2-phase} in terms of both metrics, whilst the performance gain is negligible from \textbf{2-phase} to \textbf{3-phase} in terms of $\Delta$SI-SNR. Such results implie that the 2-order latent domain constructed in \textbf{2-phase} suffices to provide a separable encoding space. Besides, the model size and GPU memory usage (on a single NVIDIA RTX 3090 when separating the 4-second speech) increases as the increases of the iterating times. These results are consistent with above theoretical analysis.}

\smallskip\noindent\textbf{Qualitative Evaluation.} To gain more insight into the effect of speech separation, we perform two sets of qualitative evaluation: 1) qualitative results by the coarse phase and refining phase and \maRevise{2) qualitative comparison between our \textbf{\emph{SRSSN}} and the \textbf{base model.}} In the first set of experiments, we randomly select two samples from the test set for DPRNN-\emph{SRSSN} and DPTNET-\emph{SRSSN}, and employ the Librosa analysis toolkit~\cite{librosa} to visualize the STFT power spectrums of the estimated speech sources in both phases in Figure~\ref{fig:real_examples}. Comparing between the visualization of the groundtruth and estimates for each involved speaker, we observe that the separated results for one speaker in the coarse phase still contain residual ingredients from the other speaker, particularly in the regions indicated by bounding boxes. 
In contrast, the separated results in the refining phase is much better than that in the coarse phase: most of the incorrect residual ingredients are successfully removed. 
The estimates in the refining phase show more similar spectrum patterns as their groundtruth counterparts than that in the coarse phase. It indicates that the fine-grained embedding space defined by the high-order latent domain in refining phase enables a more precise separation.

\maRevise{In the second  set of qualitative experiments, we randomly select two samples from the subset of \textit{High similarity} (the most challenging subset) and compare between our \textbf{\emph{SRSSN}} and \textbf{Base model} qualitatively. Figure~\ref{fig:real_examples_diff} visualizes the STFT power spectrums of the estimated speech sources. It is clearly shown that our \textbf{\emph{SRSSN}} is able to perform a more precise speech separation than the base model.}



\begin{table}[!t]
  \centering
  \caption{Performance of different methods for speech separation on WSJ0-2mix in terms of $\Delta$SI-SNR (dB) and $\Delta$SDR (dB) in the clean setting.}
  \label{tab:comparison}
  \setlength{\tabcolsep}{2.5pt} 
  \renewcommand{\arraystretch}{1.4}
  \resizebox{1\linewidth}{!}{
  \begin{tabular}
  {llccc}
  \toprule
  & Method & Model size & $\Delta$SI-SNR $\uparrow$ & $\Delta$SDR $\uparrow$ \\
  \midrule
  \multirow{4}{*}{\shortstack{Frequency \\domain-based}} & DPCL++~\cite{DPCL2}  & 13.6M & 10.8 & $-$ \\
  & UPIT-Bi-LSTM-ST~\cite{UPIT} & 92.7M & $-$ & 10.0 \\
  & Chimera++~\cite{Chimera++} & 32.9M & 11.5 & 12.0 \\
  & Deep CASA~\cite{CASA} & 12.8M & 17.7 & 18.0 \\
  \midrule
  \multirow{12}{*}{\shortstack{Learnable latent \\domain-based}} & Bi-LSTM-TASNET~\cite{BLSTM-TasNet} & 23.6M & 13.2 & 13.6 \\ 
  & Conv-TASNET~\cite{Conv-tasnet}  & 5.1M & 15.3 & 15.6 \\
  & E2EPF~\cite{post} & $-$ & 16.9 & 17.3 \\
  & FurcaNeXt~\cite{furcanext} & 51.4M & $-$ & 18.4 \\
  & DPRNN-TASNET~\cite{DPRNN} & 2.6M & 18.8 & 19.0 \\
  & SuDoRM-RF~\cite{sudo} & 2.6M & 18.9 & $-$ \\
  & Nachmani et al.~\cite{MULCAT} & 7.5M & 20.1 & $-$ \\
  & DPTNET-TASNET~\cite{DPTNet} & 2.7 M & 20.2 & 20.6 \\
  & SepFormer~\cite{SepFormer} & 26M & 20.4 & 20.5 \\
  & Wavesplit~\cite{wavesplit} & 29M & 21.0 & 21.2 \\
  \cmidrule[0.5pt]{2-5}
  & DPRNN-\emph{SRSSN} (ours) & 7.5M & 20.5 & 20.7 \\
  & DPTNET-\emph{SRSSN} (ours) & 5.7M & \textbf{21.2} & \textbf{21.4} \\
  \midrule
& \maRevise{Wavesplit $+$ \textit{Data Augment}~\cite{wavesplit}} & 29M & 22.2 & 22.3 \\
  & \maRevise{SepFormer $+$ \textit{Data Augment}~\cite{SepFormer}} & 26M & 22.3 & 22.4 \\
  \bottomrule
  \end{tabular} }
\end{table}

\textit{2) Comparison with State-of-the-art Methods on WSJ0-2mix (involving 2 speakers):}
Next we conduct experiments to compare our model with state-of-the-art methods for speech separation on WSJ0-2mix dataset\cite{DPCL}. In particular, we compare our model with 2 types of methods: 1) methods performing separation in the frequency domain, including DPCL++~\cite{DPCL2}, UPIT-Bi-LSTM-ST~\cite{UPIT}, Chimera++~\cite{Chimera++} and Deep CASA~\cite{CASA}; 2) methods performing separation in a learnable latent domain 
in an end-to-end way, including Bi-LSTM-TASNET~\cite{BLSTM-TasNet}, Conv-TASNET~\cite{Conv-tasnet}, E2EPF~\cite{post},  FurcaNeXt~\cite{furcanext}, DRPNN-TASNET~\cite{DPRNN}, SuDoRM-RF~\cite{sudo}, Nachmani et al.~\cite{MULCAT}, DPTNET-TASNET~\cite{DPTNet}, SepFormer\cite{SepFormer} and Wavesplit~\cite{wavesplit}. We evaluate the performance of two versions of our \emph{SRSSN}: DPRNN-\emph{SRSSN} and DPTNET-\emph{SRSSN}. The number of blocks in both coarse separator and refining separator $R$ is set to 6.

Table~\ref{tab:comparison} presents the experimental results of different models for speech separation on WSJ0-2mix dataset~\cite{wsj0} in terms of both $\Delta$SI-SNR and $\Delta$SDR. 
\maRevise{For the results without using data augmentation with dynamic mixing~\cite{wavesplit},} our DPRNN-\emph{SRSSN} outperforms all other methods except Wavesplit~\cite{wavesplit} in terms of both metrics while DPTNET-\emph{SRSSN} performs better than all other methods, which demonstrates advantages of our model. The methods which learn a separable encoding space defined by a latent domain, generally perform better than the other type of methods separating speech in frequency domain explicitly, which implies that the frequency domain is not necessarily the best separation space for speech as described in \cite{Conv-tasnet}. It is worth noting that our DPRNN-\emph{SRSSN} and DPTNET-\emph{SRSSN} outperform the original methods DPRNN-TASNET and DPTNET-TASNET by a large margin, respectively.

\begin{figure*}
  \centering
  \begin{minipage}[b]{0.48\linewidth}
    \centering
    \includegraphics[width=\textwidth]{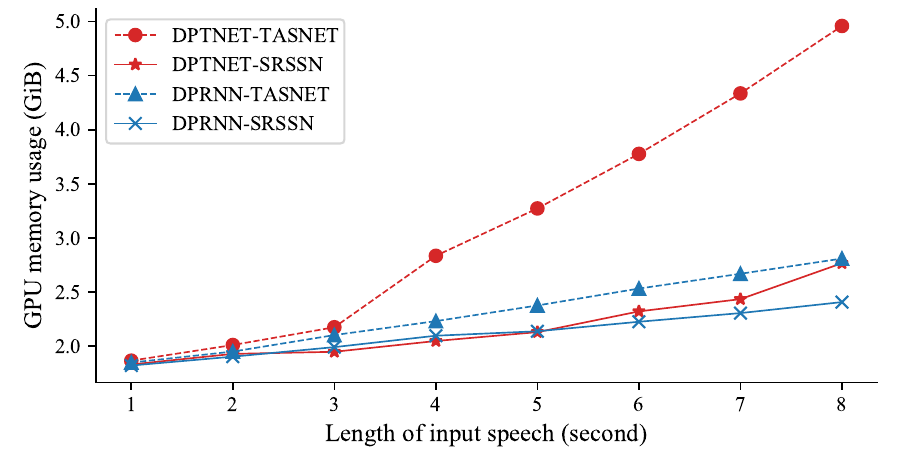}
    \centerline{(a) Comparison of GPU memory usage.}\medskip
  \end{minipage}
  \hspace{0.02\linewidth}
  \begin{minipage}[b]{0.48\linewidth}
    \centering
    \includegraphics[width=\textwidth]{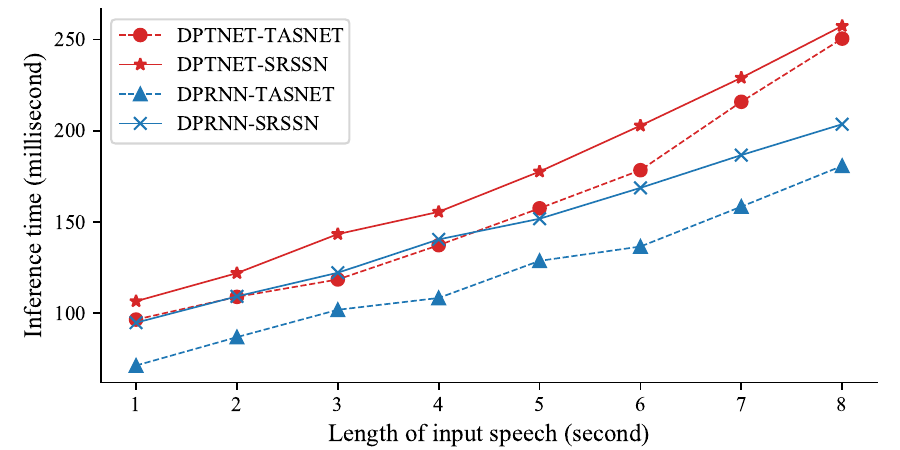}
    \centerline{(b) \maRevise{Comparison of inference time.}}\medskip
  \end{minipage}
  \caption{Comparison of GPU memory usage \maRevise{and inference time} as a function of input speech length at 8kHz sampling rate. The results are reported in inference mode on a single NVIDIA RTX 3090.}
  \label{fig:memory_time}
\end{figure*}

SepFormer~\cite{SepFormer}, which is extended over DPTNET~\cite{DPTNet} by using deeper and wider Transformer layers, achieves better performance than DPTNET at the cost of much larger model size. Nachmani et al.~\cite{MULCAT} conducts an additional task to minimize the distance of the learned speaker embeddings between the estimates and the groundtruth, which is extracted by a speaker recognition model. Such extra supervision further improves the separation performance whilst it relies on extra datasets to train the speaker recognition model. Thus it utilizes extra data for training than our model. Wavesplit~\cite{wavesplit} learns speaker-discriminative vectors at each time step by leveraging the information of speaker identities in datasets to boost performance in the training process. Compare to Nachmani et al.~\cite{MULCAT}, SepFormer\cite{SepFormer}, and Wavesplit~\cite{wavesplit}, our method DPTNET-\emph{SRSSN} achieves more superior performance without using additional information in a lightweight way. The techniques used in these models can be readily integrated into our \emph{SRSSN}, leading to a more powerful speech separation system. 

We also compare the memory consumption \maRevise{and inference time} between two versions of our model (DPRNN-\emph{SRSSN} and DPTNET-\emph{SRSSN}) with their TASNET-based counterparts (DPRNN-TASNET and DPTNET-TASNET) using the same separator structure. Figure~\ref{fig:memory_time} presents the comparison results in inference mode on the same GPU (a single NVIDIA RTX 3090) of four models when separating mixed speech of different speech duration (in second).
DPRNN-\emph{SRSSN} consumes slightly less memory than DPRNN-TASNET~\cite{DPRNN}, while DPTNET-\emph{SRSSN} reduces much more memory usage compared to DPTNET-TASNET~\cite{DPTNet}, as the length of input speech increases.
The major decrease of memory consumption lies in the larger stride size of encoder in our models than the TASNET-based methods. As reported in \cite{DPRNN}, smaller stride size of encoder leads to better performance. 
However, smaller stride size results in longer feature sequence, which requires more floating-point operations and memory usage. The stride size of encoder is tuned to be 1 in DPRNN-TASNET~\cite{DPRNN} and DPTNET-TASNET\cite{DPTNet} for optimal performance, while it is set to 8 in our coarse encoder. Our method achieves much better performance with a larger stride than TASNET-based methods, which is favorable for devices with limited memory.

\maRevise{Regarding the inference time, DPRNN-\emph{SRSSN} and DPTNET-\emph{SRSSN} both require slightly more inference time than their TASNET-based counterparts due to two separation phases in our \emph{SRSSN}}.




\begin{table}[!t]
  \centering
  \caption{\maRevise{Performance of different methods for speech separation on WSJ0-3mix in terms of $\Delta$SI-SNR (dB) and $\Delta$SDR (dB) in the clean setting.}}
  \label{tab:comparison_3spk}
  \setlength{\tabcolsep}{2.5pt} 
  \renewcommand{\arraystretch}{1.4}
  \resizebox{1\linewidth}{!}{
  \begin{tabular}
  {llccc}
  \toprule
  & Method & Model size & $\Delta$SI-SNR $\uparrow$ & $\Delta$SDR $\uparrow$ \\
  \midrule
  \multirow{2}{*}{\shortstack{Frequency \\domain-based}} & DPCL++~\cite{DPCL2}  & 13.6M & 7.1 & $-$ \\
  & UPIT-Bi-LSTM-ST~\cite{UPIT} & 92.7M & $-$ & 7.7 \\
  \midrule
  \multirow{11}{*}{\shortstack{Learnable latent \\domain-based}} & E2EPF~\cite{post} & $-$ & 12.5 & 13.0 \\
  & Conv-TASNET~\cite{Conv-tasnet} & 5.1M & 12.7 & 13.1 \\
  & DPRNN-TASNET~\cite{DPRNN} & 2.6M & 15.7 & 16.0 \\
  & DPTNET-TASNET~\cite{DPTNet} & 2.7 M & 16.2 & 16.5 \\
  & Nachmani et al.~\cite{MULCAT} & 7.5M & 16.9 & $-$ \\
  & Wavesplit~\cite{wavesplit} & 29M & 17.3 & 17.6 \\
  & SepFormer~\cite{SepFormer} & 26M & 17.6 & 17.9 \\
  \cmidrule[0.5pt]{2-5}
  & DPRNN-\emph{SRSSN} (ours) & 7.5M & 18.8 & 19.0 \\
  & DPTNET-\emph{SRSSN} (ours) & 5.7M & \textbf{19.4} & \textbf{19.6} \\
  \midrule
& Wavesplit $+$ \textit{Data Augment}~\cite{wavesplit} & 29M & 17.8 & 18.1 \\
  & SepFormer $+$ \textit{Data Augment}~\cite{SepFormer} & 26M & 19.5 & 19.7 \\
  \bottomrule
  \end{tabular} }
\end{table}

\maRevise{
\textit{3) Comparison with State-of-the-art Methods on WSJ0-3mix (involving 3 speakers):}
Next we conduct experiments to compare our model with state-of-the-art methods for speech separation on WSJ0-3mix dataset\cite{DPCL} involving 3 speakers, which is more challenging than the scenario with 2 speakers. In particular, we compare our model with 2 types of methods: 1) methods performing separation in the frequency domain, including DPCL++~\cite{DPCL2}, UPIT-Bi-LSTM-ST~\cite{UPIT}; 2) methods performing separation in a learnable latent domain, including E2EPF~\cite{post}, Conv-TASNET~\cite{Conv-tasnet}, DRPNN-TASNET~\cite{DPRNN}, DPTNET-TASNET~\cite{DPTNet}, Nachmani et al.~\cite{MULCAT}, Wavesplit~\cite{wavesplit}, and SepFormer\cite{SepFormer}. We evaluate the performance of two versions of our \emph{SRSSN}: DPRNN-\emph{SRSSN} and DPTNET-\emph{SRSSN}. The number of blocks in both coarse separator and refining separator $R$ is set to 6. 
}

\maRevise{
Table~\ref{tab:comparison_3spk} presents the experimental results of different models for speech separation on WSJ0-3mix dataset~\cite{DPCL} in terms of both $\Delta$SI-SNR and $\Delta$SDR. Our DPRNN-\emph{SRSSN} and DPTNET-\emph{SRSSN} both outperform all other methods without using data augmentation by a large margin. DPTNET-\emph{SRSSN} performs slightly worse than SepFormer using data augmentation~\cite{SepFormer}, which manifests the significant advantages of our \emph{SRSSN}. It is worth noting that our DPRNN-\emph{SRSSN} and DPTNET-\emph{SRSSN} both outperform their original methods DPRNN-TASNET and DPTNET-TASNET significantly.
}

\begin{table}[!t]
  \centering
  \caption{Performance of Speech Separation by different methods in terms of $\Delta$SI-SNR (dB) and $\Delta$SDR (dB) on WHAM! and WHAMR! in noisy and reverberant settings.}
  \label{tab:comparison_noisy}
  \setlength{\tabcolsep}{2pt} 
  \renewcommand{\arraystretch}{1.4}
  \resizebox{1\linewidth}{!}{
  \begin{tabular}
  {lcccc}
  \toprule
  \multirow{2}{*}{Method} & \multicolumn{2}{c}{WHAM!} & \multicolumn{2}{c}{WHAMR!} \\
  & $\Delta$SI-SNR $\uparrow$ & $\Delta$SDR $\uparrow$ & $\Delta$SI-SNR $\uparrow$ & $\Delta$SDR $\uparrow$ \\
  \midrule
  Chimera++~\cite{WHAM} & 9.9 & $-$ & $-$ & $-$ \\
  Bi-LSTM-TASNET~\cite{WHAMR} & 12.0 & $-$ & 9.2 & $-$ \\
  Conv-TASNET~\cite{filterbank, WHAMR} & 12.7 & $-$ & 8.3 & $-$ \\
  Learnable fbank~\cite{filterbank} & 12.9 & $-$ & $-$ & $-$ \\
  Cascaded-Bi-LSTM-TASNET~\cite{WHAMR} &  12.9 & $-$ & 10.8 & $-$ \\
  DPRNN-TASNET~\cite{MULCAT} & 13.9 & $-$ & 10.3 & $-$ \\
  DPTNET-TASNET & 14.9 & 15.3 & 12.1 & 11.1 \\
  Nachmani et al.~\cite{MULCAT} & 15.2 & $-$ & 12.2 & $-$ \\
  Wavesplit~\cite{wavesplit} & 15.4 & 15.8 & 12.0 & 11.1 \\
  \midrule
  DPRNN-\emph{SRSSN} (ours) & 15.7 & 16.1 & \textbf{12.3} & \textbf{11.4} \\
  DPTNET-\emph{SRSSN} (ours) & \textbf{16.1} & \textbf{16.5} & \textbf{12.3} & 11.3 \\
  \midrule
   \maRevise{Wavesplit $+$ \textit{Data Augment}~\cite{wavesplit}} & 16.0 & 16.5 & 13.2 & 12.2 \\
  \bottomrule
  \end{tabular} }
\end{table}

\subsection{Speech separation in Noisy and Reverberant Settings}

In this set of experiments, we conduct experiments in noisy and reverberant settings to validate the robustness of our proposed \emph{SRSSN}.

\smallskip\noindent\textbf{Datasets.}
We perform experiments on WSJ0 Hipster Ambient Mixtures (WHAM!) dataset~\cite{WHAM} and WHAMR! dataset~\cite{WHAMR}, which are constructed based on WSJ0-2mix dataset~\cite{DPCL}. In WHAM!, each two-speaker utterance from WSJ0-2mix dataset is mixed with a unique noise sample, which is recorded in non-stationary ambient environments such as coffee shops, restaurants and bars. The random SNR value between the first (louder) speaker and the noise is sampled from a uniform distribution between $-$6 dB and $+$3 dB. To separate the clean signals for involved speakers from such noisy speech data, the models are required to perform not only speech separation but also denoising. WHAMR!~\cite{WHAMR} is an reverberant extension of WHAM!, in which synthetic reverberation noise is further fused into the input speech data. 
Thus dereverberation is also required for this data to perform thorough speech separation.

We compare our proposed DPRNN-\emph{SRSSN} and DPTNET-\emph{SRSSN} with state-of-the-art methods for speech separation in noisy and reverberant settings: Chimera++~\cite{WHAM}, Bi-LSTM-TASNET~\cite{WHAMR},   Conv-TASNET~\cite{filterbank, WHAMR}, 
Learnable fbank~\cite{filterbank}, 
Cascaded-Bi-LSTM-TASNET~\cite{WHAMR},
DPRNN-TASNET~\cite{MULCAT}, 
DPTNET-TASNET~\cite{DPTNet},
Nachmani et al.~\cite{MULCAT} and Wavesplit~\cite{wavesplit}.
Note that Cascaded-Bi-LSTM-TASNET~\cite{WHAMR} is specifically designed to adapt to the noisy and reverberant conditions. 

Table~\ref{tab:comparison_noisy} shows that our DPRNN-\emph{SRSSN} and DPTNET-\emph{SRSSN} outperform other methods without using data augmentation by a large margin under noisy and reverberant conditions.
In particular, our method performs distinctly better than the cascaded model Cascaded-Bi-LSTM-TASNET~\cite{WHAMR}, which is equipped with the denoising and dereverberation functions. It manifests our model is generalized well to the noisy and reverberant conditions without specific design for adaptation.

\begin{table}[t]
  \centering
  \caption{Performance of speech recognition as well as speech separation on Libri2Mix dataset.}
  \label{tab:comparison_libri}
  \renewcommand{\arraystretch}{1.4}
  \resizebox{1.0\linewidth}{!}{
  \begin{tabular}
  {lccc}
  \toprule
  Method & $\Delta$SI-SNR (dB) $\uparrow$ & $\Delta$SDR (dB) $\uparrow$ & WER (\%) $\downarrow$\\
  \midrule
  Bi-LSTM-TASNET & 13.5 & 13.9 & 30.8 \\ 
  Conv-TASNET  & 14.4 & 14.7 & 27.4 \\
  DPRNN-TASNET & 16.1 & 16.6 & 23.8\\
  DPTNET-TASNET & 16.7 & 17.1 & 22.4 \\
  \midrule
  DPRNN-\emph{SRSSN (ours)} & 17.3 & 17.7 & 22.1 \\
  DPTNET-\emph{SRSSN} (ours) & \textbf{18.3} & \textbf{18.6} & \textbf{20.6} \\
  \midrule
  \textit{Target signal} & $-$ & $-$ & 15.6 \\
  \textit{Mixed signal} & $-$ & $-$ & 95.7 \\
  \bottomrule
  \end{tabular} }
\end{table}

\subsection{Speech Recognition on Separated speech}

We further conduct experiments of speech recognition on separated speech signals decoded by methods for speech separation to evaluate the performance of speech separation indirectly.  
To be specific, we first perform speech separation on a mixture of speech dataset, then we perform speech recognition using a standard Automatic Speech Recognition (ASR) model on the separated speech signals by different speech separation models, respectively. The achieved performance of speech recognition on the separation results can be considered as an indirect evaluation measurement for the corresponding model for speech separation.

\noindent\textbf{Dataset} 
We conduct experiments on a recently released and fully open-source dataset Libri2Mix~\cite{librimix} for speech recognition. Libri2Mix is generated based on the ASR dataset LibriSpeech~\cite{librispeech} by mixing randomly selected speech utterances from different speakers. 
We use the speech data in the clean condition with sampling rate of 8kHz. It consists of two modes \textit{min} and \textit{max}. In the \textit{min} mode, the longer utterance is trimmed to align the shorter utterance. In the \textit{max} mode, the shorter utterance is padded with zeros to align the longer utterance. 
We train the models for speech separation in the \textit{min} mode of the \textit{train-100} subset, and perform test of speech separation in the \textit{max} mode. 
The separated signals in the test phase are further used for performing experiments of speech recognition.

We use the standard hybrid DNN-HMM framework~\cite{DNN-HMM} as the ASR model to perform speech recognition, implemented based on Kaldi open-source toolkit~\cite{kaldi}. The DNN acoustic model with p-norm non-linearities~\cite{pnorm} is trained on the top of fMLLR features and the forced alignment of the training data is produced by a GMM-HMM model~\cite{compact}. A 4-gram language model is utilized for rescoring. The ASR model is trained on the subset \textit{train-clean-100} of LirbiSpeech~\cite{librispeech}, following the official kaldi implementation \footnote[1]{https://github.com/kaldi-asr/kaldi/tree/master/egs/librispeech}. We compare our proposed DPRNN-\emph{SRSSN} and DPTNET-\emph{SRSSN} with following state-of-the-art models for speech separation: Bi-LSTM-TASNET~\cite{BLSTM-TasNet}, Conv-TASNET~\cite{Conv-tasnet}, DPRNN-TASNET~\cite{DPRNN}, DPTNET-TASNET~\cite{DPTNet}. 

Table~\ref{tab:comparison_libri} presents the performance of speech recognition. Besides, we also the report the experimental results of speech separation. \textit{Target signal} and \textit{Mixed signal} denote the ASR results on the target signals and original mixed (unseparated) signals respectively, which can be viewed as the upper bound and the lower bound for the speech recognition by the same ASR model. 
Our model achieves the best performance in both speech recognition and speech separation. Besides, it is shown that the performance of speech recognition is consistent with the performance of speech separation: better speech separated results lead to higher performance of speech recognition, which reveals the effectiveness of such indirect evaluation way, namely performing speech recognition on the separated speech signals.
\label{sec:experiments}

\section{Conclusion}

In this work, we have presented the Stepwise-Refining Speech Separation Network (\emph{SRSSN}), which performs speech separation following a coarse-to-fine framework. 
\emph{SRSSN} first conducts a rough speech separation by learning a 1-order latent domain to define the encoding space in the coarse phase, then performs refining in the constructed fine-grained embedding space to achieve more precise separation. In particular, we propose the Fine-grained Encoding Mechanism, which learns a new latent domain along each basis function of the existing latent domain that defines the coarse embedding space. Thus two latent domains jointly form a high-order domain and thereby define a fine-grained embedding space. Extensive experiments have demonstrated the effectiveness of the proposed \emph{SRSSN}.

\label{sec:conclusion}

\bibliographystyle{IEEEtran}
\bibliography{refs} 

%

\end{document}